# Assaults on Judicial Independence under the Pretense of Modernization: Evidence from Venezuela[1]

Nuno Garoupa        Virginia Rosales        Rok Spruk

## Abstract

*We investigate how government-orchestrated assaults on the judiciary, disguised as modernization efforts, undermine judicial independence. Our study focuses on Venezuela's constitutional overhaul in the early 2000s, initiated by Hugo Chávez and implemented through a judicial emergency committee. We employ a hybrid synthetic control and difference-in-differences approach to estimate the impact of populist attacks on judicial independence trajectories. By comparing Venezuela to a stable pool of countries without radical constitutional changes, our identification strategy isolates the effect of populist assaults from unobservable confounders and common time trends. Our findings reveal that authoritarian interventions lead to an immediate and lasting breakdown of judicial independence. The deterioration in judicial independence vis-á-vis the estimated counterfactual is robust to variations in the donor pool composition. It does not appear to be driven by pre-existing judicial changes and withstands numerous temporal and spatial placebo checks across over nine million randomly sequenced donor samples.*

[1] Garoupa: Professor of Law, Antonin Scalia Law School, George Mason University, 3301 Fairfax Drive, Arlington, VA 22201. E: ngaroup@gmu.edu. Rosales: Professor of Political Economy, Department of Applied Economics, University of Granada, Campus Cartuja, Spain. E: vrosales@ugr.es. Spruk (corresponding author): Associate Professor of Economics, School of Economics and Business, University of Ljubljana, Kardeljeva ploscad 17, SI-1000 Ljubljana. Research Fellow, Business School, University of Western Australia, 8716 Hackett Dr., Crawley WA 6009, Australia. E: rok.spruk@ef.uni-lj.si. We are grateful to the editor, Justin McCrary, one anonymous referee, William Hubbard, Keith Hylton, Anna Lewczuk, and 2025 World Congress of Econometric Society (Seoul), ALEA 2024 (Ann Arbor), ESELS 2024 (Elche), AEDE 2024 (Barcelona), and EALE 2024 (Torino), University of Granada and Tel Aviv University seminar participants for helpful comments. The usual disclaimers apply.

# 1 Introduction

Few would dispute that a well-functioning and independent judiciary is essential for sustainable governance and economic development. As such, the evaluation of legal system reforms, and their consequences for judicial independence, has attracted significant scholarly attention. Among the methodological innovations in this area, the synthetic control method has emerged as one of the most powerful tools for policy evaluation, praised for its rigor and transparency (Athey and Imbens 2006; Xu 2017; Abadie and Cattaneo 2018; Abadie 2021; Chernozhukov et al. 2021; Pang et al. 2022). Described as one of the most important contributions to empirical inference in the past two decades (Athey and Imbens 2017), the method aims to credibly estimate counterfactual trajectories. That is, how outcomes would have evolved in the absence of a specific intervention, by reconstructing and replicating pre-intervention outcome trajectories using an optimally weighted control group.

The institutional literature on judicial independence is well established (Ginsburg 2003; Howard and Carey 2004; Helmke and Rosenbluth 2009; Epperly and Sievert 2019; Dijk 2021), and a growing body of work highlights the rising frequency of assaults on judiciaries worldwide (Voeten 2020). This article investigates how such assaults, carried out under the rhetorical guise of judicial modernization, affect the integrity and independence of courts. We focus on a critical early case of the constitutional transformation initiated by Hugo Chávez during the onset of Venezuela's Bolivarian revolution. Far from simply reforming the judiciary, the Chávez administration used a battery of legal instruments, especially the rule of judicial emergency committee, to purge judges, pack the Supreme Court with loyalists, and systematically dismantle the judiciary's ability to constrain executive power.

To isolate the causal effects of this authoritarian intervention, we compare Venezuela to a carefully constructed donor pool of Ibero-American countries that were not subject to analogous institutional disruptions. This comparative framework enables us to determine whether the observed collapse in judicial independence stems specifically from the constitutional transformation, rather than from broader regional dynamics or macroeconomic decline. Our empirical strategy integrates difference-in-differences and synthetic control techniques, applied to annual data spanning from 1960 to 2021. This hybrid approach affords both intuitive causal inference and rigorous counterfactual modeling.

The empirical evidence indicates that the 1999 constitutional overhaul precipitated a rapid and persistent deterioration in judicial independence. The decline is particularly acute in measures capturing de facto high-court autonomy, compliance with judicial decisions, constraints on executive influence, and mechanisms of judicial accountability. Our findings suggest that strategic court-packing and politically motivated appointments were central instruments of institutional degradation. Far from representing a transient episode, these changes became entrenched within the judiciary's structure, eroding its autonomy and enabling systemic corruption and executive interference in judicial outcomes.

To ensure the robustness of these results, we employ a suite of advanced estimators—including classical difference-in-differences, synthetic control, synthetic difference-in-differences, matrix completion, and interactive fixed effects, and conduct both temporal and spatial placebo tests. Across this comprehensive battery of empirical strategies, the estimated negative effects of the authoritarian intervention remain consistent, irrespective of donor pool composition or model specification. This methodological robustness underscores the profound and enduring consequences of authoritarian judicial restructuring, revealing its capacity to fundamentally alter the institutional fabric of the judiciary and compromise the very foundations of the rule of law.

Finally, our findings speak directly to ongoing debates in comparative politics about the timing and nature of institutional decay in Venezuela (García Holgado and Sánchez Urribarri 2024; Polga-Hecimovich and Sánchez Urribarri 2025). While much attention has focused on the 2004 reforms and later developments under Nicolás Maduro, our analysis indicates that the most severe institutional damage occurred early, under Hugo Chávez. The results support the interpretation that authoritarian legalism represents a foundational structural break, not a gradual continuum, with subsequent reforms merely entrenching rather than initiating the collapse of judicial autonomy.

The remainder of the paper proceeds as follows. Section 2 reviews the relevant literature and provides historical context for the Venezuelan case. Section 3 outlines the identification strategy. Section 4 describes the data and sample construction. Section 5 presents the results and robustness checks. Section 6 concludes.

## 2    Literature Review and Institutional Context

The literature on authoritarianism has paid attention to courts for some time (Ginsburg and Moustafa 2008). Once in power, authoritarian governments continuously undermine both judicial institutions and civil liberties as part of the effort to promote an alternative governance model to a well-established liberal democracy (Ginsburg et. al. 2018). In many contexts, as in Venezuela, such policy is pursued under the pretext of modernization.

There is by now plenty of literature on the rise of Chávez and the expansion of authoritarian Bolivarianism in Venezuela and outside (de la Torre 2017, Corrales 2023) as well as the unfolding collapse of democratic institutions (Canache 2002, Morgan 2007, Hsieh et. al. 2011). The changes in the political regime in Venezuela are more characterized as transitions from democracy to competitive authoritarianism (Mainwaring 2012), not simply left-wing populism or democratic backsliding (Mudde and Kaltwasser 2017, Urbinati 2019, Tushnet and Bugaric 2021). In his work, Corrales (2015) described the Venezuelan political regime as autocratic legalism, that is, limited rule of law, where the authoritarian policies are subject to legality, also called rule by law, and the courts play a relevant role in protecting the executive from possible legal contestation.

Three main events took place in Venezuela since the election of President Hugo Chávez in 1998. First, in 1999, the National Assembly declared an emergency within the Venezuelan judicial system, creating the "Emergency Judicial Commission" with the power to evaluate judges' performance. This new program was presented as a form of modernizing a corrupt and inefficient judiciary. A 2000 Enabling Law followed, thus concentrating power on the executive. Second, in 2004, the Organic Law of the Supreme Court was enacted which concluded the reform of the judicial apex institutions by packing with loyalists. Third, after President Nicolás Maduro took over in 2013, new judicial reforms were enacted to guarantee a submissive judiciary, including a reform of the Organic Law of the Supreme Court in 2022. In Table A1, in appendix, we summarize the main law reforms related to the Venezuelan judicial system and implemented from 1999, when Hugo Chávez convened the constitutional referendum, to 2022, when the last reform of the Organic Law of the Supreme Court was enacted.

Different scholars have identified various key dates to mark the structural break in judicial independence and quality in Venezuela. An earlier group of scholars, including Garcia-Serra (2001), Castaldi (2006), and Brewer-Carias (2009, 2010), emphasized the significance of the 2000 Enabling Law. This reform drastically altered the separation of powers, effectively eliminating judicial independence and paving the way for an authoritarian government committed to ruling by law rather than upholding the rule of law. Since 2000, the executive branch has maintained tight control over judicial appointments, and a 2004 law further entrenched this control by packing the Supreme Court with loyalists. All justices became subject to presidential oversight, including the possibility of dismissal through a vote in the executive-controlled legislature. As a result, both the judiciary and legislature were subordinated to an increasingly dominant executive power following several institutional reforms. Furthermore, Rodriguez (2022) noted that Chávez exploited his constitutional majoritarian power to undermine counter-majoritarian institutions, such as the judiciary, thereby accelerating the rise of authoritarianism and the erosion of checks and balances.

More broadly, Garoupa and Maldonado (2011) argued that Chávez's earlier judicial reforms were facilitated by the judiciary's already diminished credibility by 1999, which was plagued by corruption scandals, inefficiency, significant case backlogs, weak political influence, and excessive legalism. From an earlier period, institutional capture in Venezuela ensured that regime supporters were deeply entrenched in the judiciary. For instance, Sánchez Urribarri (2011, 2024) and Taylor (2014) described how the Supreme Court became an instrument of authoritarianism after Chávez systematically eliminated any opposition from within the judiciary.

Recent scholarship by Bennaim (2020) has focused on the significant impact of the 2004 law, highlighting three major consequences: the imposition of limited judicial tenure (where judges could be dismissed if they displeased the regime), the purging and packing of the judiciary, and the subordination of judicial decisions to political interference. Bennaim also points to 2015-2017 as a critical moment when the Supreme Court abandoned any pretense of independence and began to legitimize unconstitutional actions by the president, marking a further deterioration of judicial autonomy (a point further debated by Polga-Hecimovich and Sánchez Urribarri 2025).

The existing literature in comparative constitutional law indicates a strong association between the rise of populism and decline of the rule of law (Lacey 2019, Gora and De Wilde 2022) relying on short-term correlations between populism and deterioration of indicators used to capture judicial independence. Yet, the causal mechanism behind the correlations remains subject to rigorous debate and cannot be established by simple correlation (Lewkowicz et. al. 2024). In a previous article, Garoupa and Spruk (2024a) introduced more advanced econometric techniques to assess the impact of court reforms on indicators of judicial quality, with an application to Türkiye.

## 3 Identification Strategy

### 3.1 Contextual setup

The key goal of our identification strategy is to examine the contribution of populist constitutional reforms undertaken by Chávez administration in 1999 to judicial quality and independence consistently. The general objective of our strategy hinges on the estimation of the missing counterfactual scenario associated with the 1999 constitutional overhaul and the subsequent erosion of judicial independence. In this respect, our empirical strategy is to estimate the series of judicial independence trajectories in the hypothetical absence of the populist constitutional assaults under the reform pretext. In comparison with a more popular and traditional difference-in-differences setup, our key variable of interest is the trajectory of judicial independence up to the present day rather than merely average treatment effect itself. The validity of the difference-in-differences approach hinges on the presence of parallel trends between the affected units and their control peers. In the permissible absence of parallel trends assumption, the validity of the average treatment effect is rendered questionable which may jeopardize both the plausibility and interpretability of the estimated effects. Contrary to the difference-in-differences estimator, synthetic control method does not necessitate the presence of parallel trends as the control groups are built through a two-stage training and validation period where outcome- and covariate-level similarities are properly ascribed (Abadie et. al. 2010, 2015, Abadie 2021, Gilchrist et. al. 2023). A more detailed setup of the counterfactual design is provided in Supplementary Appendix A. Furthermore, Supplementary Appendices B and C elaborate extensively on the in-space and in-time placebo tests that we carry out to test the statistical significance of the estimated gaps.

## 4  Data and Samples

### 4.1  Outcome variables

We collect data on judicial independence by compiling a variety of indicators using the recently updated Varieties of Democracy dataset (Lindberg et. al. 2014, Coppedge et. al., 2016, 2022). The dataset collates more than 480 indicators and transforms into five core indices along with other supplementary indices including the independence of judiciary. Six different indicators of judicial independence are considered as the outcomes of interest in our investigation on an annual basis and included in the vector of dependent variables. The variables have been created from large-scale surveys of country-level experts and were transformed into the ordinal scale using the Bayesian item response theory measurement model. We will refer to these six indicators as judicial independence, although different scholars can quibble with the nature of one or more indicators. For example, Aydin-Cakir (2023) focused exclusively on high court independence in her study about Hungary and Poland. Since the empirical results are presented for each indicator individually, one can leave to the reader the decision about which indicators are more or less compelling as appropriate and relevant measures of de facto judicial independence.

First, high-court independence variable reflects the independence and autonomy of the judiciary in the decision-making process without the interference of fear of the executive branch of government. Thus, the variable can identify the autonomous judicial decision making and its absence. It can be compared across space and time. By default, judicial decisions can reflect government preferences and wishes as a court can have a high degree of autonomy whilst its decisions support the government's position. Judges can also be persuaded about the merits of the government's position. The variable reflects whether the court simply adopts the government adoption with no regard to the sincere view of the government record.[2] Second, lower court independence variable reflects the degree of judicial independence in courts below the Supreme Court level. In particular, the variable is reflective of when judges not on the high court are ruling in cases that are salient to the government, how often and likely their decisions merely reflect government wishes regardless of their sincere and professional view of the legal record.[3]

Third, court packing variable indicates the degree to which the executive branch of government promulgates politically motivated judicial appointments to the high and lower court. The size of the judiciary is at times increased for certain reasons such as the increasing caseload or purely for political reasons when the executive branch aims to influence the judiciary.[4] Fourth, compliance with high court variable reflects the degree to which the executive branch of government complies with important decisions of the high court with which it disagrees.[5] Fifth, judicial constraints on the executive reflect and summarize the overall strength of the judiciary in constraining arbitrary government action that oftentimes violates the constitution and existing legislation.

---

[2] The variable can take five distinctive values: (i) zero if the court always adopts the government's position, (ii) one if the high court usually adopts the government's position, (iii) two if the high court adopts the government's position about one half of the time, (iv) three if the high court rarely adopts the government's position, and (v) four if the court never adopts the government's position. Higher values of the variable thereby indicate a somewhat greater and more resilient degree of judicial independence.

[3] More specifically, the variable is ordinal and can take five possible values (i) zero if the lower court decisions always reflect the government's position, (ii) one if the lower courts usually adopt the government's position, (iii) two if the lower court adopts the government's position about one half of the time, (iv) three if the lower court rarely adopts the government's position, and (v) four if the court never adopts the government's position. Higher values of the variable thereby indicate a somewhat greater and more resilient degree of judicial independence.

[4] The variable can take four distinctive values: (i) 3 if judgeships were added to the judiciary, but there is no evidence that the increase was politically-motivated and there was simply no increase, (ii) 2 if judgeships were added to the judiciary and there was a limited politically-motivated increase in the number of judgeships, (iii) 1 if there was a limited, politically-motivated increase in the number of judgeships on very important courts, and (iv) 0 if there was a massive, politically-motivated increase in the number of judgeships across the entire judiciary. Therefore, lower values of the variables on the ordinal scale indicate a greater interference and active involvement of the executive branch of government in the judicial appointments.

[5] The variable can take five distinctive values that have been converted to the interval using item response measurement model: (i) zero if the government never complies, (ii) one if the government seldom complies, (iii) two if the government complies about half of the time, (iv) three if the government usually complies, and (v) four if the government always complies with the important decisions of the high court. Higher values simply indicate a higher overall degree of compliance and respect of the judicial autonomy and decisions of the high court.

This particular variable is a high-level indicator of the extent to which the executive branch of government respects the constitution, complies with court rulings and the extent to which the judiciary can act independently of government pressure or threats.[6]

Sixth, judicial purges variable indicates the degree to which judges are removed from their posts for arbitrary and typically political reasons, and not for reasons such as strong evidence of corruption and abuse of judicial power.[7] Seventh, judicial decision corruption variable indicates how often do individuals or businesses make undocumented extra payments or bribes to speed up or delay the process to obtain a favorable judicial decision.[8]

Eighth, government attacks on judiciary variable reflects how often the government attacked the integrity of the judiciary in public. Attacks on the integrity of judiciary can be performed through claims that it is corrupt, incompetent or that decisions were politically motivated. These assaults can manifest in a variety of ways such as prepared statements reported by the media, press conferences, interviews and stump speeches.[9] And ninth, judicial accountability variable indicates the degree to which judges are removed from their posts or otherwise disciplined when found responsible for serious misconduct.[10]

## 4.2    Sample

---

[6] In terms of further detail, the index is constructed through a point estimate from Bayesian factor analysis model of the following sub-level indicators: (i) respect of the constitution by executive branch, (ii) compliance with judiciary, (iii) compliance with high court, (iv) high-court independence, (v) lower-court independence. Higher values in the interval designate more rigorous and stronger judicial constraints on the executive branch of government.

[7] The variable can take five distinctive values: (i) zero if there was a massive and arbitrary purge of the judiciary, (ii) one if there were limited but very important arbitrary removals, (iii) two if there were limited arbitrary removals, (iv) three if judges were removed from office but there is no evidence that removals were arbitrary, and (v) four if judges were not removed from their posts. The variable is measured on the ordinal scale and converted to the interval through the item-response measurement model.

[8] The variable is ordinal and can take five distinctive values, specifically (i) zero if extra payments or bribes are always paid, (ii) one if usually paid, (iii) two if bribes and extra payments are paid about half of the time, (iv) three if this is not a usual practice, and (v) four if bribery and extra payments never happen. The variable is converted to the interval using Bayesian item response theory measurement model.

[9] In this respect, the variable can take five different values (i) zero if attacks were carried out on a daily or weekly basis, (ii) one if assaults were common and carried out in nearly every month of the year, (iii) if attacks occurred more than once, (iv) three if attacks occurred but were rate, and (v) if there were no attacks on the integrity of the judiciary. Hence, higher values of the variable imply fewer government assaults on the judiciary at lower frequency. The variable is ordinal and is converted to the interval based on the item response model in a cross-coder aggregation.

[10] The variable can take five different values: (i) zero if removal and disciplinary procedure never happen, (ii) one if the removal and disciplinary procedure seldom happen, (iii) two if judges are removed from their post or disciplined about half of the time, (iv) three if removal and disciplinary action usually happen, and (v) four if removal and disciplinary action always happen. Higher values of the variable indicate a substantially greater degree of judicial accountability.

Our period of investigation begins in 1960 and ends in 2021. By early 1960s, Venezuela has accomplished both political and economic liberalization and transitioned to the democratic rule which provides a relatively stable and enduring time period without major overhauls and civil unrest that would pose a source of systemic instability. The overall sample comprises Venezuela as a single-treated country and a donor pool of 17 Ibero-American countries[11] for the period of 62 years (i.e., 1960-2021) without modernization-disguised authoritarian subjugation of the judiciary, which yields a strongly balanced panel of 1,116 country-year paired observations. It should be noted that the necessary and perhaps fundamental condition behind the non-violation of stable treatment value assumption concerns the salience behind the identification of treatment effect. In particular, salience implies that to properly isolate the treatment effect of authoritarian assault on the judicial independence and integrity, countries undergoing authoritarian consolidation similar to Venezuela should not be included into the donor pool. In this respect, a potential threat to the identification of the treatment effect is posed by the countries whose governments have been closely aligned with the policies and agenda of Chavez government and have pursued similar subordination of the judicial branch of government. In turn, any comparison of Venezuela with such countries would preclude a reasonably credible identification of the treatment effect, and compromise its validity. Our approach to achieve salience consists of the rigorous restriction of the Ibero-American donor pool to the set of countries where the authoritarian subjugation of the judiciary was never achieved and completed. Subsequently, countries belonging to the Bolivarian Alliance for the Peoples of Our America (i.e. *Alianza Bolivariana para los Pueblos de Nuestra América*)[12], which comprise the Chavez's ring of friends (i.e. Bolivia, Cuba, Ecuador, and Nicaragua), are not included in the donor pool since their implicit attributes of the judiciary cannot be used to plausibly assess the effects of authoritarian assaults on the trajectories of judicial independence, and are subsequently discarded from the Ibero-American donor pool.

Figure 1 presents the patterns of judicial independence by plotting the means of six outcome variables over time for Venezuela and full Ibero-American donor pool. For the sake of convenience, Venezuela is highlighted dashed whilst the donor countries are highlighted in bold. Although such analysis does not per se entail any inference on the effect, it serves as an informative insight into the overall outcome dynamics before a more nuanced and sophisticated analysis is undertaken. The raw trends of judicial independence imply a graphically visible deterioration of judicial independence with respect to increasing politically motivated judicial appointments, lesser degree of independence and compliance, weak constraints and higher intensity of judicial purges after the populist assault on the judiciary instigated in 1999.

---

[11] Argentina, Brazil, Chile, Colombia, Costa Rica, Dominican Republic, El Salvador, Equatorial Guinea, Guatemala, Honduras, Mexico, Nicaragua, Panama, Paraguay, Peru, Portugal, Spain, and Uruguay.

[12] *Alianza Bolivariana para los Pueblos de Nuestra América* (ALBA) is intergovernmental organization supporting the political and economic integration of Latin American and Caribbean countries. It was founded in 2004 by Cuba and Venezuela and is associated with socialist and social democratic governments wishing to consolidated regional economic integration based on the premises of social welfare and mutual economic assistance.

**Figure 1**: Plotting the trends of judicial independence in Venezuela and the donor pool

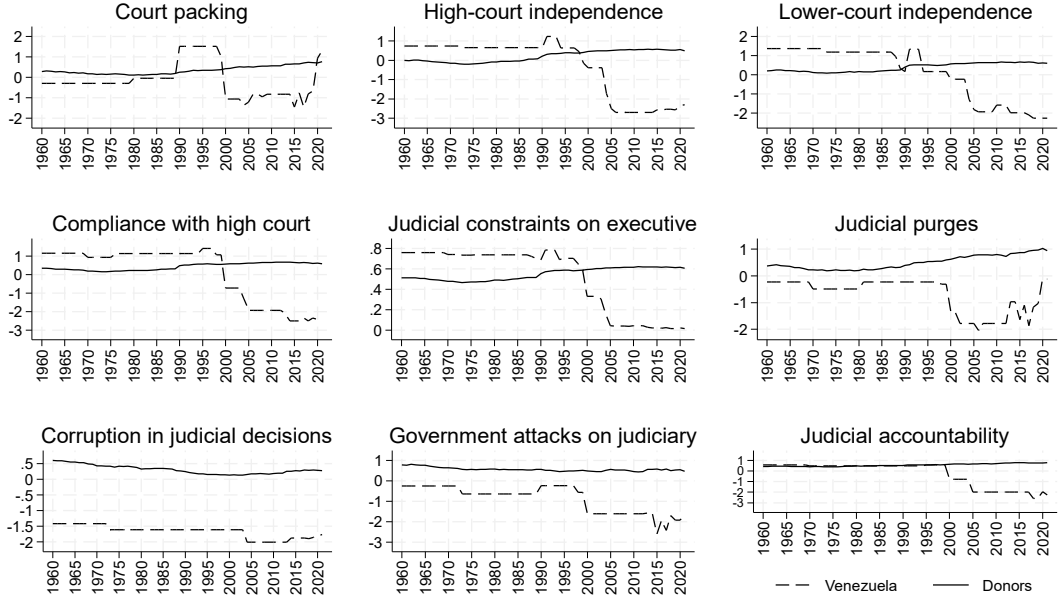

Table 1 presents the balance of pre-overhaul judicial independence outcome path between Venezuela and its synthetic peers using constrained nested optimization route and full-outcome path matching algorithm. The evidence from pre-overhaul matching of outcome values invariably suggests that the synthetic control estimator provides an excellent quality of the fit since the outcome values of Venezuela are almost identical to the corresponding peers in outcome-specific synthetic control groups. A reasonably high quality of fit implies two imminent features. First, the trajectories of Venezuela's judicial independence in the pre-overhaul period can be well approximated from the convex attributes of judicial independence among its peers in the donor pool. And second, a nearly exact similarity between pre-$T_0$ outcome path of Venezuela and its synthetic peers also implies that both systemic and idiosyncratic shocks intrinsic to Venezuela's judiciary are unlikely to exert a major confounding influence per se. If such shocks were pivotal, the actual and synthetic trajectories of judicial independence would relinquish a divergent pattern unlikely to lend a feasible credence to the internal validity and stability of the estimated gaps.

**Table 1** Balancing pre-assault judicial independence paths

| | Court packing | | High-court independence | | Lower-court independence | | Compliance with high court | | Judicial executive constraints | | Judicial purges | | Judicial corruption | | Government attacks on judiciary | | Judicial accountability | |
|---|---|---|---|---|---|---|---|---|---|---|---|---|---|---|---|---|---|---|
| | Real | Synthetic | Real | Synthetic | Real | Synthetic | Real | Synthetic | Real | Synthetic | Real | Synthetic | Real | Synthetic | Real | Synthetic | Real | Synthetic |
| RMSE | 0.005 | | 0.041 | | 0.184 | | 0.026 | | 0.039 | | 0.047 | | 0.039 | | 0.009 | | 0.009 | |
| 1960 | -0.294 | -0.293 | 0.747 | 0.751 | 1.369 | 1.042 | 1.153 | 1.162 | 0.758 | 0.744 | -0.227 | -0.223 | -1.421 | -1.362 | -0.260 | -0.263 | 0.575 | 0.574 |
| 1961 | -0.294 | -0.293 | 0.747 | 0.751 | 1.369 | 1.346 | 1.153 | 1.162 | 0.758 | 0.760 | -0.227 | -0.224 | -1.421 | -1.362 | -0.260 | -0.264 | 0.575 | 0.574 |
| 1962 | -0.294 | -0.293 | 0.747 | 0.744 | 1.369 | 1.346 | 1.153 | 1.162 | 0.758 | 0.758 | -0.227 | -0.212 | -1.421 | -1.396 | -0.260 | -0.248 | 0.575 | 0.575 |
| 1963 | -0.294 | -0.293 | 0.747 | 0.746 | 1.369 | 1.346 | 1.153 | 1.152 | 0.758 | 0.758 | -0.227 | -0.229 | -1.421 | -1.396 | -0.260 | -0.262 | 0.575 | 0.575 |
| 1964 | -0.294 | -0.293 | 0.747 | 0.746 | 1.369 | 1.357 | 1.153 | 1.164 | 0.758 | 0.759 | -0.227 | -0.223 | -1.421 | -1.396 | -0.260 | -0.261 | 0.575 | 0.574 |
| 1965 | -0.294 | -0.293 | 0.747 | 0.744 | 1.369 | 1.357 | 1.153 | 1.209 | 0.758 | 0.760 | -0.227 | -0.225 | -1.421 | -1.396 | -0.260 | -0.271 | 0.575 | 0.574 |
| 1966 | -0.294 | -0.293 | 0.747 | 0.744 | 1.369 | 1.336 | 1.153 | 1.120 | 0.758 | 0.760 | -0.227 | -0.223 | -1.421 | -1.396 | -0.260 | -0.256 | 0.575 | 0.574 |
| 1967 | -0.294 | -0.293 | 0.747 | 0.752 | 1.369 | 1.336 | 1.153 | 1.122 | 0.758 | 0.749 | -0.227 | -0.220 | -1.421 | -1.446 | -0.260 | -0.244 | 0.575 | 0.574 |
| 1968 | -0.294 | -0.293 | 0.747 | 0.738 | 1.369 | 1.329 | 1.153 | 1.137 | 0.758 | 0.747 | -0.227 | -0.233 | -1.421 | -1.446 | -0.260 | -0.253 | 0.575 | 0.574 |
| 1969 | -0.294 | -0.294 | 0.747 | 0.738 | 1.369 | 1.322 | 0.928 | 0.939 | 0.740 | 0.739 | -0.486 | -0.473 | -1.421 | -1.464 | -0.260 | -0.264 | 0.464 | 0.463 |
| 1970 | -0.294 | -0.294 | 0.747 | 0.738 | 1.369 | 1.321 | 0.928 | 0.939 | 0.740 | 0.740 | -0.486 | -0.486 | -1.421 | -1.464 | -0.260 | -0.266 | 0.464 | 0.463 |
| 1971 | -0.294 | -0.294 | 0.747 | 0.725 | 1.369 | 1.244 | 0.928 | 0.938 | 0.740 | 0.737 | -0.486 | -0.474 | -1.421 | -1.538 | -0.260 | -0.275 | 0.464 | 0.463 |
| 1972 | -0.294 | -0.305 | 0.664 | 0.720 | 1.176 | 1.244 | 0.928 | 0.938 | 0.735 | 0.736 | -0.486 | -0.484 | -1.611 | -1.546 | -0.637 | -0.625 | 0.464 | 0.463 |
| 1973 | -0.294 | -0.284 | 0.664 | 0.629 | 1.176 | 1.244 | 0.928 | 0.946 | 0.735 | 0.736 | -0.486 | -0.481 | -1.611 | -1.591 | -0.637 | -0.661 | 0.464 | 0.463 |
| 1974 | -0.294 | -0.284 | 0.664 | 0.661 | 1.176 | 1.244 | 0.928 | 0.992 | 0.735 | 0.734 | -0.486 | -0.484 | -1.611 | -1.591 | -0.637 | -0.645 | 0.464 | 0.463 |
| 1975 | -0.294 | -0.284 | 0.664 | 0.661 | 1.176 | 1.244 | 1.141 | 1.117 | 0.738 | 0.738 | -0.486 | -0.481 | -1.611 | -1.591 | -0.637 | -0.645 | 0.464 | 0.463 |
| 1976 | -0.294 | -0.284 | 0.664 | 0.661 | 1.176 | 1.244 | 1.141 | 1.121 | 0.738 | 0.738 | -0.486 | -0.484 | -1.611 | -1.591 | -0.637 | -0.626 | 0.464 | 0.463 |
| 1977 | -0.294 | -0.295 | 0.664 | 0.661 | 1.176 | 1.168 | 1.141 | 1.138 | 0.738 | 0.737 | -0.486 | -0.483 | -1.611 | -1.591 | -0.637 | -0.627 | 0.464 | 0.463 |
| 1978 | -0.294 | -0.295 | 0.664 | 0.664 | 1.176 | 1.159 | 1.141 | 1.140 | 0.738 | 0.738 | -0.486 | -0.478 | -1.611 | -1.606 | -0.637 | -0.639 | 0.464 | 0.463 |
| 1979 | -0.294 | -0.295 | 0.664 | 0.665 | 1.176 | 1.159 | 1.141 | 1.146 | 0.738 | 0.734 | -0.225 | -0.222 | -1.611 | -1.594 | -0.637 | -0.646 | 0.464 | 0.463 |
| 1980 | -0.054 | -0.055 | 0.664 | 0.652 | 1.176 | 1.156 | 1.141 | 1.146 | 0.738 | 0.730 | -0.225 | -0.233 | -1.611 | -1.594 | -0.637 | -0.638 | 0.464 | 0.463 |
| 1981 | -0.054 | -0.055 | 0.664 | 0.665 | 1.176 | 1.156 | 1.141 | 1.146 | 0.738 | 0.735 | -0.225 | -0.219 | -1.611 | -1.592 | -0.637 | -0.636 | 0.464 | 0.463 |
| 1982 | -0.054 | -0.055 | 0.664 | 0.665 | 1.176 | 1.156 | 1.141 | 1.146 | 0.738 | 0.735 | -0.225 | -0.220 | -1.611 | -1.623 | -0.637 | -0.642 | 0.464 | 0.463 |
| 1983 | -0.054 | -0.055 | 0.664 | 0.666 | 1.176 | 1.163 | 1.141 | 1.147 | 0.738 | 0.737 | -0.225 | -0.226 | -1.611 | -1.642 | -0.637 | -0.645 | 0.464 | 0.463 |
| 1984 | -0.054 | -0.055 | 0.664 | 0.666 | 1.176 | 1.163 | 1.141 | 1.147 | 0.738 | 0.737 | -0.225 | -0.224 | -1.611 | -1.659 | -0.637 | -0.630 | 0.464 | 0.463 |
| 1985 | -0.054 | -0.055 | 0.664 | 0.645 | 1.176 | 1.163 | 1.141 | 1.148 | 0.738 | 0.737 | -0.225 | -0.221 | -1.611 | -1.587 | -0.637 | -0.637 | 0.464 | 0.462 |
| 1986 | -0.054 | -0.055 | 0.664 | 0.645 | 0.942 | 0.751 | 1.141 | 1.148 | 0.721 | 0.711 | -0.225 | -0.223 | -1.611 | -1.589 | -0.637 | -0.645 | 0.464 | 0.462 |
| 1987 | -0.054 | -0.055 | 0.664 | 0.683 | 0.343 | 0.734 | 1.141 | 1.147 | 0.701 | 0.707 | -0.225 | -0.222 | -1.611 | -1.589 | -0.637 | -0.622 | 0.464 | 0.462 |
| 1988 | -0.054 | -0.055 | 0.664 | 0.683 | 0.171 | 0.731 | 1.141 | 1.144 | 0.685 | 0.709 | -0.225 | -0.226 | -1.611 | -1.631 | -0.243 | -0.253 | 0.464 | 0.462 |
| 1989 | -0.054 | -0.026 | 0.664 | 0.673 | 1.340 | 0.956 | 1.141 | 1.143 | 0.785 | 0.775 | -0.225 | -0.231 | -1.611 | -1.560 | -0.243 | -0.253 | 0.464 | 0.462 |
| 1990 | 1.514 | 1.512 | 0.664 | 0.677 | 1.340 | 0.956 | 1.141 | 1.139 | 0.785 | 0.773 | -0.225 | -0.213 | -1.611 | -1.560 | -0.243 | -0.244 | 0.464 | 0.462 |
| 1991 | 1.514 | 1.512 | 1.243 | 1.233 | 1.340 | 0.956 | 1.141 | 1.133 | 0.785 | 0.778 | -0.225 | -0.225 | -1.611 | -1.566 | -0.243 | -0.242 | 0.464 | 0.463 |
| 1992 | 1.514 | 1.512 | 1.243 | 1.192 | 0.175 | 0.577 | 1.141 | 1.233 | 0.690 | 0.710 | -0.225 | -0.222 | -1.611 | -1.571 | -0.243 | -0.246 | 0.464 | 0.463 |
| 1993 | 1.514 | 1.512 | 1.243 | 1.170 | 0.175 | 0.411 | 1.414 | 1.379 | 0.702 | 0.694 | -0.225 | -0.222 | -1.611 | -1.546 | -0.243 | -0.251 | 0.581 | 0.580 |
| 1994 | 1.514 | 1.512 | 0.653 | 0.821 | 0.175 | 0.197 | 0.702 | 0.739 | 0.702 | 0.692 | -0.225 | -0.217 | -1.611 | -1.663 | -0.243 | -0.245 | 0.581 | 0.580 |
| 1995 | 1.514 | 1.512 | 0.653 | 0.617 | 0.175 | 0.204 | 1.414 | 1.379 | 0.702 | 0.678 | -0.225 | -0.235 | -1.611 | -1.663 | -0.243 | -0.245 | 0.581 | 0.580 |
| 1996 | 1.514 | 1.512 | 0.653 | 0.623 | 0.175 | 0.205 | 1.067 | 1.112 | 0.651 | 0.676 | -0.310 | -0.302 | -1.611 | -1.663 | -0.560 | -0.547 | 0.581 | 0.580 |
| 1997 | 1.514 | 1.512 | 0.653 | 0.547 | 1.369 | 1.042 | 1.153 | 1.162 | 0.758 | 0.744 | -0.227 | -0.223 | -1.421 | -1.362 | -0.260 | -0.263 | 0.575 | 0.574 |
| 1998 | 1.514 | 1.512 | 0.445 | 0.546 | 1.369 | 1.346 | 1.153 | 1.162 | 0.758 | 0.760 | -0.227 | -0.224 | -1.421 | -1.362 | -0.260 | -0.264 | 0.575 | 0.574 |

### 4.3 Measurement Error and Threats to External Validity

A common concern in empirical research on institutional quality relates to the use of perception-based indicators derived from expert surveys, which are frequently criticized for their inherent subjectivity, opaque sampling methodologies, and potential coder biases. Such limitations raise the possibility that these measures may amplify measurement error and introduce substantial noise, potentially obscuring the underlying signal of judicial independence across countries and over time. In light of these concerns, a central question arises: do alternative indices of judicial independence—proposed and widely debated in the existing literature—capture patterns comparable to those reflected in the expert perception-based indicator employed in our analysis? To address this issue and partially mitigate the limitations associated with reliance on a single source, we undertake a systematic comparison of our core indicators of judicial independence and integrity against a set of alternative longitudinal measures, namely, World Justice Project (2021), World Bank's rule of law index (Kaufmann et. al. 2011), Freedom House index of the rule of law (Freedom House 2024), Linzer and Staton (2015) judicial independence index and Cingranelli and Richards (2010) judicial independence score.[13]

Table 2 presents the correlation matrix between the Varieties of Democracy (Coppedge et. al. 2022) indices of judicial independence and integrity versus the respective alternative indices. The comparison reveals a high and robust degree of longitudinal concordance between the indicators employed in our empirical analysis and a set of alternative measures, thereby providing compelling evidence that the longitudinal indicators used are neither high-leverage outliers nor unduly noisy. Rather, the measures of judicial independence adopted in our study closely mirror the broader empirical trajectories identified by leading datasets in the literature, reinforcing their validity. To further address concerns regarding reliance on a single measure, the matrix includes p-values (reported in parentheses) for tests of the null hypothesis that the correlation coefficient equals zero. The outcome variables in our analysis exhibit moderate to strong and statistically significant correlations with the full array of longitudinal indicators. Notably, the strongest and most statistically significant correlations are observed for indicators pertaining to compliance with high-court rulings, high-court independence, judicial constraints on the executive, judicial corruption, and judicial accountability.

---

[13] A growing debate has questioned whether current indicators, particularly Varieties of Democracy, accurately capture democratic backsliding in Europe and North America since the 2010s. While V-Dem is often favored over alternatives like Polity IV and Freedom House, scholars remain divided over its conceptual foundations, coding biases, and interpretations of long-term institutional decline (Garoupa and Spruk 2024b).

**Table 2**: Cross-index correlation matrix of judicial independence metrics

|  | World Justice Project | WGI Rule of Law (Kaufmann et. al. 2011) | Freedom House Rule of Law index | Linzer-Staton Judicial Independence Index | Cingrangelli and Richards Judicial Independence Score |
|---|---|---|---|---|---|
| Court packing | +0.59 | +0.57 | +0.62 | +0.63 | +0.46 |
|  | (0.000) | (0.000) | (0.000) | (0.000) | (0.000) |
| High-court independence | +0.71 | +0.68 | +0.75 | +0.68 | +0.47 |
|  | (0.000) | (0.000) | (0.000) | (0.000) | (0.000) |
| Lower-court independence | +0.79 | +0.68 | +0.76 | +0.68 | +0.55 |
|  | (0.000) | (0.000) | (0.000) | (0.000) | (0.000) |
| Compliance with high court | +0.85 | +0.71 | +0.79 | +0.78 | +0.58 |
|  | (0.000) | (0.000) | (0.000) | (0.000) | (0.000) |
| Judicial constraints on executive | +0.81 | +0.73 | +0.82 | +0.82 | +0.59 |
|  | (0.000) | (0.000) | (0.000) | (0.000) | (0.000) |
| Judicial purges | +0.60 | +0.69 | +0.69 | +0.64 | +0.43 |
|  | (0.000) | (0.000) | (0.000) | (0.000) | (0.000) |
| Judicial corruption | +0.91 | +0.86 | +0.75 | +0.62 | +0.59 |
|  | (0.000) | (0.000) | (0.000) | (0.000) | (0.000) |
| Government attacks on judiciary | +0.62 | +0.52 | +0.37 | +0.43 | +0.55 |
|  | (0.000) | (0.000) | (0.000) | (0.000) | (0.000) |
| Judicial accountability | +0.84 | +0.72 | +0.61 | +0.79 | +0.56 |
|  | (0.000) | (0.000) | (0.000) | (0.000) | (0.000) |

To further assuage the caveats behind single-measure dependence, Figure 2 presents the correlation between the weighted average of our nine indicators, obtained as a first principal component, against the World Justice Project and Kaufmann et. al. (2011) indices of the rule law as well as Linzer and Staton (2015) judicial independence score. The comparison reveals a very high correlation in the range between +0.82 (i.e. p-value = 0.000), and +0.86 (i.e. p-value = 0.000) and confirms the central finding of the judicial independence dynamics being both persistent and uniform regardless of the specific indicator. Across all three specific comparison, Venezuela's score of judicial independence and rule of law is consistently ranked worst in our full sample. Whilst it is impossible to fully eliminate subjectivity from any such index, the convergence of scores across diverse data sources bolsters our confidence that the observed patterns of judicial independence dynamics reflect substantive institutional degradation rather than spurious artefacts and measurement error. It is at least illustrative of the external validity of our results (Besley and Persson 2011, Hollyer et. al. 2015). Nonetheless, it should be stressed and acknowledged that all currently available indicators of judicial independence and rule of law entail some level of imprecision that is rather a question of degree than of kind. Therefore, we refrain from making any categorical claims about causality beyond the evidence supported by our identification strategy. Rather than offering definitive and broadly encompassing conclusions, our empirical analysis contributes to a growing literature that treats institutional collapses as measure and subject to rigorous empirical inquiry, however imperfect the available data may be (Acemoglu et. al. 2015)

**Figure 2**: Judicial independence and rule of law cross-index correlation in the full sample, 1960-2021

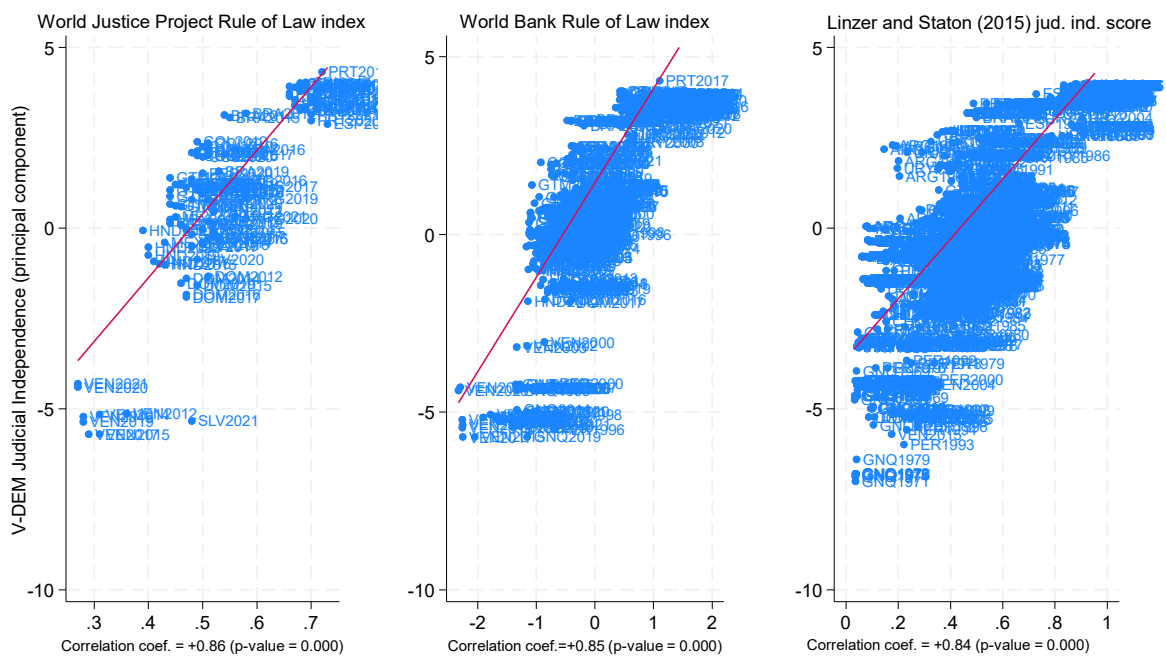

## 5  Results

### 5.1  Difference-in-differences estimates

To evaluate the institutional consequences of the 1999 authoritarian intervention—ostensibly framed as judicial modernization but widely recognized as a pivotal moment in the executive's capture of the judiciary, we employ a standard difference-in-differences (DiD) framework. This approach compares shifts in Venezuela's judicial independence trajectory before and after the intervention to those observed in a control group of countries that did not experience analogous institutional disruptions during the same period. The analysis incorporates country- and year-fixed effects to control for unobserved, time-invariant country-specific characteristics and for common temporal shocks affecting all countries in the sample. Given the singular nature of the intervention, affecting only Venezuela and occurring at a single point in time, our estimation strategy aligns with established best practices for DiD designs applied to single-unit binary treatment (Sant'Anna and Zhao 2020, Callaway and Sant'Anna 2021, Rambachan and Roth 2023).

The core identifying assumption underlying our analysis is that of parallel trends. Therefore, in the absence of the 1999 intervention, Venezuela would have exhibited judicial independence trajectories similar to those of the control countries. To strengthen the credibility of this assumption, we estimate both static and dynamic specifications of the difference-in-differences model. Across all model variants, we include controls for economic development, proxied by GDP per capita, to account for the potential confounding effects of macroeconomic decline and volatility on the integrity of institutional structures.

Table 3 reports the difference-in-differences estimates derived from a range of model specifications. Panel A presents results from the static model, which captures the average post-intervention change in Venezuela's judicial independence and integrity relative to the pre-reform period. Across all outcome variables, ranging from court-packing intensity in column (1) to judicial accountability in column (9), the estimates indicate a statistically significant and substantively pronounced decline in judicial independence following the 1999 authoritarian intervention. The deterioration is particularly acute in the case of high-court independence, for which we estimate a 71 percent decline relative to the baseline of an equally weighted pre-intervention control group (p-value = 0.000). The intervention is further associated with a marked erosion in compliance with judicial rulings, a surge in judicial purges, and the proliferation of corruption in judicial decision-making, all pointing to a sweeping consolidation of political control over the judiciary.

In Panel B, we estimate a dynamic version of the difference-in-differences specification that incorporates lagged outcome variables and country-by-year interaction terms to address potential concerns regarding unobserved, time-varying shocks in institutional trajectories. As anticipated, the magnitude of the estimated treatment effect associated with the authoritarian intervention is modestly attenuated in this richer specification, reflecting a degree of path dependence and institutional inertia that may be obscured in static models. Nevertheless, the DiD coefficients remain highly statistically significant at the 1% level and maintain consistent directional signs across all outcomes, indicating a pervasive decline in judicial independence and erosion of institutional integrity. These results further reinforce the interpretation that the 1999 authoritarian assault initiated a persistent and systemic degradation of Venezuela's judiciary.

To illustrate the temporal dynamics of this institutional collapse, Figure 3 presents an event-study plot showing year-by-year treatment effects for key indicators of judicial independence and integrity, specifically, high-court independence, compliance with high-court rulings, judicial constraints on the executive, and government-led attacks on the judiciary. The coefficients for the pre-intervention period are uniformly flat and statistically indistinguishable from zero, even under lenient significance thresholds, thereby lending strong support to the absence of anticipatory trends and affirming the credibility of the parallel trends assumption. In contrast, the post-assault coefficients reveal an immediate and sustained deterioration, consistent with the onset of a sharp structural break rather than a gradual institutional decline. The timing of disruption coincides with the establishment of the Judicial Emergency Committee and the subsequent sweeping vertical reorganization of the judiciary. Taken together, and conditional on the parallel trends assumption, our findings indicate that the deterioration in judicial quality, independence, and integrity is causally linked to the 1999 turning point. This moment marks a radical inflection in the evolution of Venezuela's judicial institutions and cannot be attributed to underlying economic conditions or broader regional trends in institutional development.

**Table 3**: Difference-in-Differences Estimates of the Effect of Authoritarian Assault on Judicial Independence in Venezuela, 1960-2021

|  | Court packing | High-court independence | Lower-court independence | Compliance with high court | Judicial constraints on executive | Judicial purges | Judicial corruption | Government attacks on judiciary | Judicial accountability |
|---|---|---|---|---|---|---|---|---|---|
|  | (1) | (2) | (3) | (4) | (5) | (6) | (7) | (8) | (9) |
| **Panel A: Two-way fixed-effects (TWFE) difference-in-differences estimates** | | | | | | | | | |
| Post-$T_0$ $\Delta Y$ | -1.721*** | -3.648*** | -2.709*** | -3.362*** | -.832*** | -1.947*** | -.450** | -1.047*** | -2.434*** |
|  | (.237) | (.330) | (.328) | (.264) | (.056) | (.333) | (.220) | (.301) | (.269) |
| Two-tailed 95% confidence bounds | [-2.222, -1.219] | [-4.344, -2.951] | [-3.402, -2.016] | [-3.921, -2.804] | [-.952, -.713] | [-2.651, -1.242] | [-.916, .014] | [-1.711, -.437] | [-3.003, -1.865] |
| Country-fixed effects | YES | YES | YES | YES | YES | YES | YES | YES | YES |
| (p-value) | (0.000) | (0.00) | (0.00) | (0.00) | (0.00) | (0.00) | (0.00) | (0.00) | (0.00) |
| Time-fixed effects | YES | YES | YES | YES | YES | YES | YES | YES | YES |
| (p-value) | (0.000) | (0.000) | (0.000) | (0.000) | (0.000) | (0.000) | (0.000) | (0.000) | (0.000) |
| Covariates | YES | YES | YES | YES | YES | YES | YES | YES | YES |
| (p-value) | (0.000) | (0.000) | (0.000) | (0.000) | (0.000) | (0.000) | (0.000) | (0.000) | (0.000) |
| **Panel B: Dynamic two-way (TWFE) difference-in-differences estimates** | | | | | | | | | |
| Post-$T_0$ $\Delta Y$ | -.448*** | -.713*** | -.398*** | -.475*** | -.126*** | -.498*** | -.032* | -.184* | -.354*** |
|  | (.093) | (.157) | (.102) | (.146) | (.024) | (.155) | (.018) | (.107) | (.053) |
| Two-tailed 95% confidence bounds | [-.647, -.250] | [-1.045, -.381] | [-.614, -.182] | [-.784, -.167] | [-.181, -.035] | [-.826, -.169] | [-.071, -.006] | [-.410, .041] | [-.467, -.241] |
| Country-Year Fixed Effects | YES | YES | YES | YES | YES | YES | YES | YES | YES |
| (p-value) | (0.000) | (0.000) | (0.000) | (0.000) | (0.000) | (0.000) | (0.000) | (0.000) | (0.000) |
| Covariates | YES | YES | YES | YES | YES | YES | YES | YES | YES |
| (p-value) | (0.000) | (0.000) | (0.000) | (0.000) | (0.000) | (0.000) | (0.000) | (0.000) | (0.000) |
| Within R2 | 0.82 | 0.87 | 0.90 | 0.91 | 0.91 | 0.77 | 0.91 | 0.74 | 0.92 |
| # observations | 1,044 | 1,044 | 1,044 | 1,044 | 1,044 | 1,044 | 1,044 | 1,044 | 1,044 |
| # countries | 18 | 18 | 18 | 18 | 18 | 18 | 18 | 18 | 18 |
| Time period | 1960-2021 | 1960-2021 | 1960-2021 | 1960-2021 | 1960-2021 | 1960-2021 | 1960-2021 | 1960-2021 | 1960-2021 |

Notes: This table reports estimates from difference-in-differences (DiD) regressions evaluating the institutional effects of the authoritarian assault on the judiciary unleashed by the 1999 judicial emergency committee rule for the period 1960-2021. All outcome specifications include country and year fixed effects to adjust the model estimates for unobserved heterogeneity bias and common time-varying judicial shocks, and the log GDP per capita to address the potential confounding role of economic downturns in shaping judicial outcomes. The identification strategy relies on the parallel trends assumption. No anticipatory effects are detected, and placebo reassignments do not yield similar patterns of institutional decline. Panel A reports results from a static two-way fixed effects (TWFE) specification with a binary treatment indicator equal to one for Venezuela in years following the 1999 assault. Panel B reports dynamic DiD estimates that allow the treatment effect to evolve over time. These specifications include lagged dependent variables and a full set of country-year interaction terms to absorb unobserved time-varying heterogeneity that may otherwise bias the DiD coefficient in a single-country treatment setting. The inclusion of lagged outcomes helps mitigate concerns about serial correlation and model misspecification, and the saturated interaction structure further reduces risk of confounding from region-specific or year-specific institutional shocks as well as the risk of model misspecification. The control group consists of all countries in our full sample that did not undergo comparable judicial reforms during the 1960-2021 estimation window. Standard errors are adjusted for heteroskedastic distribution of random error variance and serially correlated stochastic disturbances using finite-sample adjustment of the non-nested empirical distribution function based on two-fold clustering scheme at the country and year level. Standard errors are denoted in the parentheses. Asterisks denote statistically significant DiD coefficients at 10% (*), 5%(**), and 1% (***), respectively.

**Figure 3**: Effect of the Authoritarian Assaults on Judicial Independence and Integrity in Venezuela, 1960-2021

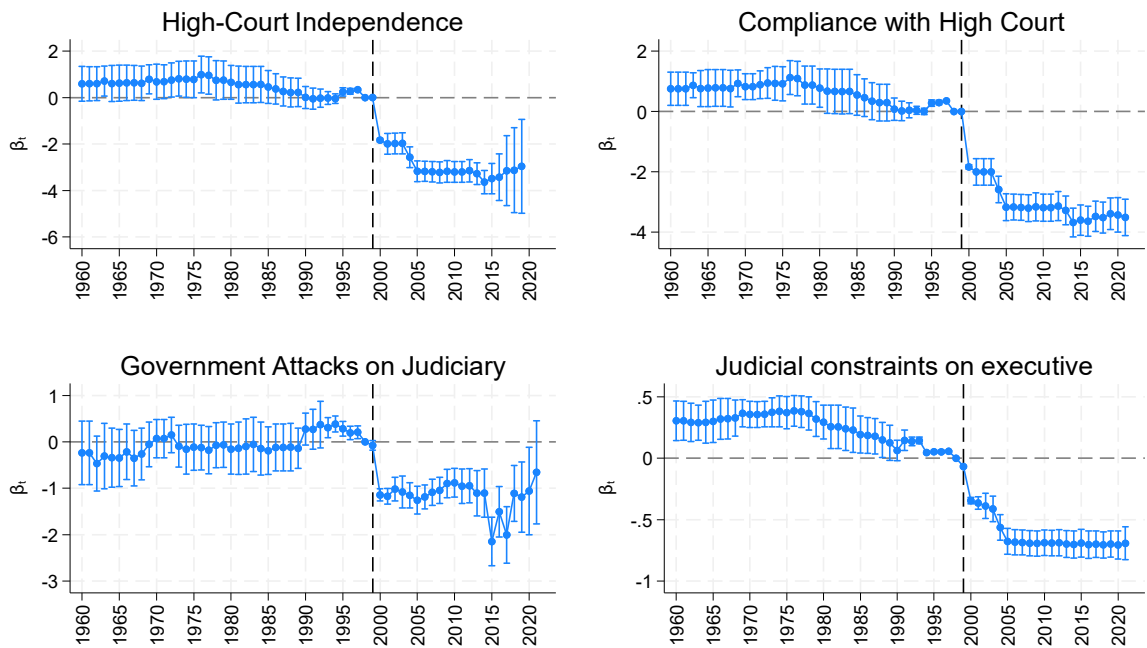

## 5.2 Synthetic control estimates

While the difference-in-differences (DiD) approach provides a transparent and widely accepted framework for estimating causal effects, it has well-documented limitations in single-unit treatment settings, particularly when the parallel trends assumption may be tenuous. Chief among these limitations is the imposition of a homogeneous treatment effect over time, as well as the strong reliance on the comparability between Venezuela's pre-assault judicial trajectory and the average trend of the donor pool. These constraints may introduce potential biases in weighting and counterfactual construction. To address these methodological concerns and allow for a more flexible, data-driven approximation of the counterfactual, we turn to the synthetic control method (SCM). This approach offers two principal advantages in our context. First, it constructs an optimally weighted combination of unaffected countries that most closely replicates Venezuela's pre-intervention path of judicial independence. Second, it eschews parametric assumptions regarding the functional form of treatment effects over time, allowing for a more nuanced estimation of dynamic institutional change.

Adopting SCM not only enhances the credibility of our identification strategy but also permits a cross-validation of the DiD results while relaxing the rigidity of its identifying assumptions. As the subsequent analysis demonstrates, the SCM results corroborate the patterns of judicial collapse identified in the DiD framework, while offering a more granular and transparent reconstruction of the counterfactual trajectory of Venezuela's judiciary in the absence of the 1999 constitutional overhaul.

Figure 4 illustrates the estimated effects of the 1999 intervention by plotting Venezuela's actual judicial independence trajectory against its synthetic counterpart for the pre- and post-intervention periods. Given the close alignment of the pre-treatment paths, the results strongly suggest that the constitutional overhaul precipitated a near-total disintegration of judicial independence. In contrast to the stable evolution of the synthetic control group, Venezuela's post-treatment trajectory exhibits a marked and enduring collapse. Specifically, the synthetic estimates indicate a dramatic and likely irreversible deterioration in judicial independence. One of the most salient indicators of this institutional decay is the substantial decline in the court-packing index. The estimates reveal a drop of approximately three points immediately following the reform, persisting until 2018, when the index begins to revert toward pre-overhaul levels. While court-packing effects may be inherently temporary, particularly when the entire composition of the judiciary is overhauled, the broader consequences of the constitutional reform remain deep and persistent.

In particular, the independence of the Supreme Court experienced a pronounced and sustained decline, with synthetic control estimates indicating an average reduction of more than 2.6 basis points in the post-intervention period. This erosion is gradual yet consistent, showing little evidence of recovery over time. In contrast, the synthetic control trajectory for Supreme Court independence remains stable throughout the same period, providing a plausible benchmark for the counterfactual scenario. A similar pattern emerges for lower-court independence, which declined by approximately 2.37 basis points. Interestingly, the gap is somewhat larger than for the high court, driven in part by the continued deterioration in Venezuela after 2015, while the synthetic counterpart shows improvement during the same period.

The constitutional overhaul also significantly undermined government compliance with Supreme Court decisions, resulting in a decline of approximately 2.6 basis points by the end of the sample period. In parallel, judicial constraints on executive power weakened substantially, with an estimated drop of 0.6 basis points. These effects are not transitory; rather, they reflect a permanent recalibration of institutional power that decisively shifted authority toward the executive. Both indicators exhibit trajectories that deviate sharply from their synthetic counterparts, reinforcing the interpretation of a systemic and lasting institutional breakdown.

While the evidence on judicial purges points to a somewhat more transient effect, the implications remain significant. The average estimated gap between Venezuela and its synthetic counterpart is 1.55 basis points. The intensity of purges increased markedly between 2000 and 2006, stabilizing thereafter until 2017, and subsequently moderating. However, this apparent decline does not necessarily indicate a substantive change in government strategy. Rather, given the near-total restructuring of the judiciary, further purges became redundant. This interpretation is supported by the continued expansion of political control, particularly through informal mechanisms of coercion and intimidation.

Consistent with this view, we observe a significant rise in judicial corruption following the overhaul. The judicial corruption index deteriorates by approximately 0.42 basis points, a particularly salient effect given that Venezuela's corruption levels increased while those of its synthetic counterpart declined, reflecting regional improvements. This widening gap underscores the extent to which the Venezuelan judiciary became a tool of political manipulation. The same trend is evident in government-led attacks on the judiciary, where our estimates indicate a deterioration of 1.47 basis points. This suggests that despite a reduction in court-packing and purging activity, overt and covert assaults on judicial institutions have not abated.

Perhaps the most compelling evidence of institutional degradation emerges from the judicial accountability indicator. Owing to the near-perfect pre-treatment fit between Venezuela and its synthetic analogue, the post-overhaul divergence is striking. Our estimates reveal an erosion of judicial accountability by approximately 2.95 basis points by the end of the sample period, suggesting a near-complete collapse of mechanisms designed to ensure judicial transparency and responsibility.

Taken together, the SCM results present a detailed and internally coherent picture of institutional decline. They not only validate the findings of the DiD framework but also expose the multifaceted and enduring consequences of the 1999 constitutional overhaul on Venezuela's judicial architecture. These findings strongly support the conclusion that the reform constituted a decisive rupture in the trajectory of judicial development—one that entrenched executive dominance and severely compromised the independence and integrity of the judiciary. Supplementary Appendix D provides a detailed composition of outcome-specific synthetic control groups. An important consideration for assessing the internal validity of our baseline findings concerns the composition of the synthetic control groups used to approximate Venezuela's judicial independence trajectories. By construction, synthetic counterparts are derived from a weighted combination of countries whose pre-treatment outcome trajectories fall within the convex hull of Venezuela's own pre-overhaul values. These weights are obtained via the convex optimization route, which identifies the set of donor countries whose outcome profiles best replicate Venezuela's institutional path prior to treatment. In Supplementary Appendix E, we perform a series of leave-one-out analyses (Klößner et. al. 2018), and show that the estimated gaps are fully robust to the exclusion of highest-leverage control countries from the donor pool.

**Figure 4**: Effects of constitutional overhaul of judicial independence in Venezuela, 1960-2021

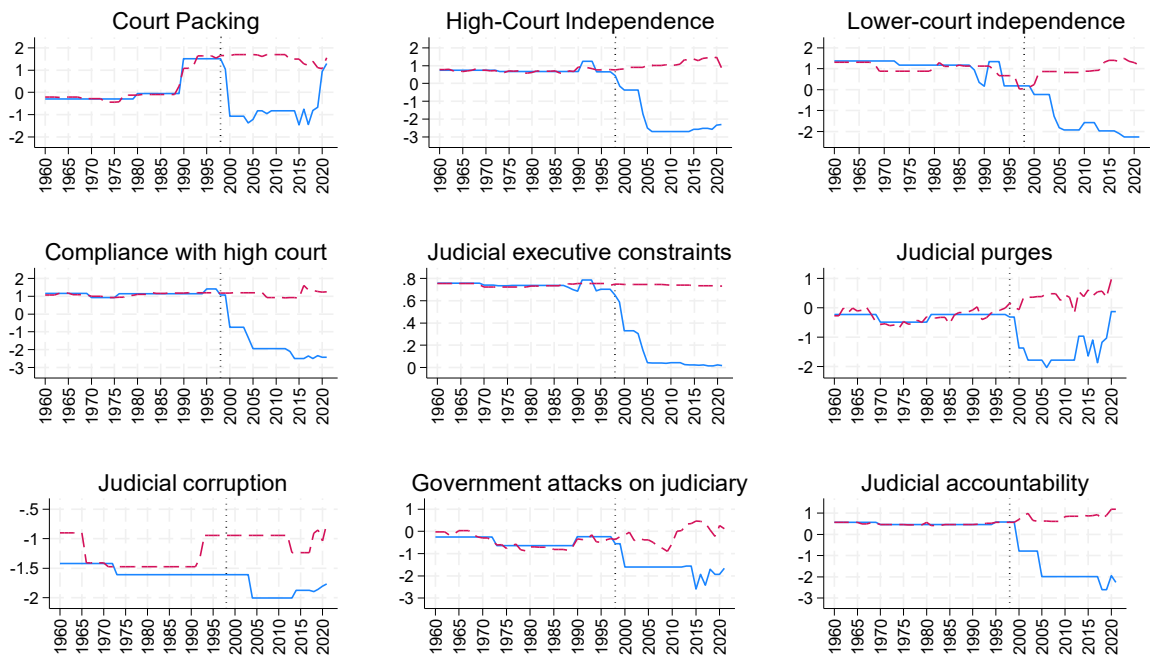

*5.3    Differential trend analysis*

An additional robustness consideration emerging from our findings concerns the dynamic properties of Venezuela's judicial independence trajectories before and after the 1999 constitutional overhaul. Without loss of generality, our synthetic control estimates reveal a sharp and accelerated erosion of judicial independence following the reform. This naturally raises a critical question: does Venezuela exhibit a statistically significant *differential trend* in judicial independence in the post-overhaul period compared to its synthetic counterpart?

If the constitutional overhaul indeed constituted a structural break, we would expect two empirical patterns to emerge: (i) statistical equivalence in the pre-overhaul period both in level (intercept) and trend (slope) between Venezuela and its synthetic control, and (ii) a statistically significant divergence in trajectory following the intervention. Following the approach of Spruk and Kovac (2020), we formally test for a trend break by conducting a triple-differences analysis, comparing pre- and post-treatment changes in Venezuela's outcomes to those of its synthetic twin. This method embeds the classical Chow (1960) structural break framework within an exact Fisher testing environment to evaluate whether the treatment induced a significant change in trend. We compute two-tailed test statistics and corresponding p-values to assess the null hypothesis of no differential trend, and construct 95% confidence intervals and standard errors to gauge the extent and precision of the post-treatment shift.

The differential trend framework offers notable advantages. It allows us to detect not only whether the judicial independence gap widened following the reform, but also whether the constitutional intervention altered the *rate* at which institutional deterioration unfolded. Failure to reject the null hypothesis would suggest continuity in pre-existing trends, whereas rejection implies that the reform triggered a distinct and more severe trajectory, thereby indicating evidence of a deeper causal mechanism beyond mere correlation. In this way, testing for differential trends helps distinguish between weak and strong forms of assault on judiciary.

Collectively, the evidence provides strong empirical support for the interpretation that the 1999 constitutional overhaul marked not only a disruption in institutional levels, but also a structural reorientation in judicial trajectories. Table 4 reports the corresponding Supremum Wald test statistics for each outcome, confirming the presence of significant differential trends and thus substantiating the claim of an authoritarian rupture with enduring institutional consequences.

**Table 5**: Testing differential trend assumption behind the judicial independence effect of constitutional overhaul

|  | Court packing | High-court independence | Lower-court independence | Compliance with high court | Judicial constraints on executive | Judicial purges | Judicial corruption | Government attacks on judiciary | Judicial accountability |
|---|---|---|---|---|---|---|---|---|---|
|  | (1) | (2) | (3) | (4) | (5) | (6) | (7) | (8) | (9) |
| Panel A: Wald-Chow structural break test with unknown break point |
| Supremum Wald $\chi^2$ test statistics (p-value) | 4143.57 (0.000) | 285.19 (0.000) | 17.58 (0.003) | 35.58 (0.000) | 38.28 (0.000) | 328.33 (0.000) | 201.71 (0.000) | 49.43 (0.000) | 891.90 (0.000) |
| Panel B: Wald-Chow structural break test known break point (year = 1999) |
| Supremum Wald $\chi^2$ test statistics (p-value) | 21.48 (0.000) | 37.17 (0.000) | 6.36 (0.041) | 14.57 (0.000) | 29.76 (0.000) | 328.33 (0.000) | 7.02 (0.029) | 44.64 (0.000) | 891.90 (0.000) |

Notes: the table reports *Spruk and Kovac (2020)* test of the differential trend assumption behind the estimated judicial independence gaps. Under the null hypothesis, the constitutional overhaul fails to produce a differential trend of the judicial independence gap between Venezuela and its synthetic control group. Under the null hypothesis, both Venezuela and its synthetic control group exhibit similar judicial independence gap trend between pre- and post-overhaul period. The table reports the p-values on the exact Supremum-Wald test statistics of the Chow (1960) structural break in the triple-differences structural setup of the model. Two-sided p-values are reported in the parentheses.

*5.6    Additional robustness checks*

To bolster the validity of our estimates further, we carry an additional battery of robustness checks, namely, generalized synthetic control analysis (Xu 2017), reported in Supplementary Appendix F, analysis based on donor pool replacement with Mercosur member states, reported in Supplementary Appendix G, LASSO synthetic control analysis under non-convex optimization and countercyclical weights (Hollingsworth and Wing 2020), reported in Supplementary Appendix H, and recently proposed analysis based on synthetic difference-in-differences estimator (Arkhangelsky et. al. 2021), with more flexible unit-level additive shifts, where results are reported extensively in Supplementary Appendix I. The estimated impact of the constitutional overhaul on judicial independence appears to be fully consistent with our baseline results, and robust to the choice of treatment effect estimator.

# 6    Conclusion

This paper has examined how authoritarian interventions, cloaked in the rhetoric of judicial modernization, have fundamentally undermined judicial independence in Venezuela. Focusing on the 1999 constitutional overhaul and the subsequent establishment of a judicial emergency committee, we employ a hybrid empirical strategy, combining synthetic control and difference-in-differences methods, drawing on the Varieties of Democracy dataset and a carefully curated donor pool of 17 Ibero-American countries that did not undergo comparable institutional disruptions. Our counterfactual analysis reveals a pronounced and durable erosion of judicial integrity following the overhaul, placing Venezuela on a sharply divergent institutional trajectory relative to its regional peers.

The empirical evidence suggests that this authoritarian transformation produced an immediate and sustained breakdown of judicial independence, marked by politically motivated judicial appointments, the dismantling of compliance with the Supreme Court, the systematic weakening of executive constraints, a proliferation of judicial corruption, and the collapse of accountability mechanisms. These findings are robust across multiple outcome measures, donor pool compositions, and estimation techniques—including interactive fixed effects, matrix completion, and synthetic difference-in-differences estimators. The effects are statistically significant and not attributable to pre-treatment imbalances or latent institutional trends.

To further validate these results, we conduct rigorous falsification tests through placebo interventions across unaffected countries and placebo years derived from data-driven structural break tests. These exercises confirm that the estimated declines in judicial independence are neither artifacts of spurious shocks nor temporary deviations, but rather reflect a structural inflection point in Venezuela's institutional development. Subsequent reforms in 2004 and 2015 appear to reinforce, rather than reverse, the trajectory set in motion by the 1999 overhaul.

Taken together, our findings underscore the transformative and lasting consequences of authoritarian judicial restructuring. They demonstrate that institutional degradation can be swift, deliberate, and difficult to reverse, even in systems already perceived as fragile. While the erosion of judicial independence in Venezuela may seem unsurprising in retrospect, our analysis provides a rare empirical account of just how far and how fast authoritarian legal engineering can dismantle institutional checks. The fact that the post-1999 breakdown was not inevitable but contingent, measurable, and distinct from regional trends, offers a sobering reminder: however compromised an institution may appear, it can always get worse.

Beyond the Venezuelan case, this study offers a methodological template for assessing the causal impact of constitutional interventions in similarly unstable or transitional regimes. By refining the empirical toolkit available for studying court-packing, judicial capture, and constitutional backsliding, we contribute to a growing literature on the mechanics of authoritarian consolidation and the structural vulnerabilities of legal systems. Our approach is applicable across contexts where democratic erosion occurs under the guise of legality and institutional reform.

We intentionally refrain from drawing inferences regarding downstream economic or political outcomes, as such extensions require distinct identification strategies, alternative data structures, and longer post-treatment observation windows than those available in the present study. While the judicial system undoubtedly interacts with a wide array of economic and political institutions, establishing causal links between judicial erosion and broader macro-political outcomes would necessitate addressing additional confounders, dynamic feedback effects, and potentially endogenous institutional responses. As such, we deliberately limit the scope of our analysis to the institutional domain of judicial independence, where our identification strategy is most credible and empirically well-grounded.

Consistent with Klick's (2010, 2018) cautionary guidance on causal inference in institutional contexts, our research design prioritizes transparency, methodological rigor, and robustness to high-dimensional covariates that often characterize complex governance systems. In doing so, this study establishes a new empirical benchmark for the causal analysis of court-packing, judicial capture, and authoritarian constitutional transformation. By leveraging advanced synthetic control techniques, including flexible extrapolation methods and hybrid estimators, we offer a replicable and theoretically informed framework for evaluating institutional decay. This framework lays the groundwork for future research exploring the broader consequences of judicial degradation, including its potential effects on democratic resilience, state capacity, and the long-term stability of political regimes.

# Supplementary Appendix - A Brief Summary of Legal Reforms in Venezuela

**Table A1:** The "Judicial Revolution" in Venezuela: main reforms (1999-2022)

| Year | Description |
|---|---|
| 1999 (April 25) | President Hugo Chávez, elected in 1998, convened a constituent referendum, proposing the election of a National Assembly to write a new constitution for Venezuela. Chávez won the constituent referendum with 88% of the votes in favor. This marked the beginning of the constituent process in the country (Pérez-Perdomo, 2007). |
| 1999 (July 25) | Chávez convened elections to vote for members of the National Assembly. From a total of 131 seats, 125 were pro-government supporters, and 6 were opposition supporters. |
| 1999 (August 18) | The National Assembly declared an emergency within the Venezuelan Judicial System, creating the "Emergency Judicial Commission" with the power to evaluate judges' performance. According to authors such as Pérez-Perdomo (2007) and Louza Scognamiglio (2017), between 1999-2003 there was a judicial purge in Venezuela (340 judges were impeached and dismissed). |
| 1999 (December 15) | The National Assembly finished the constituent process. Chávez convened a referendum to validate the new constitution. There were several deep changes related to the former constitution, especially concerning the branches of government (executive, legislative, and judicial). In the case of the judiciary, the *Corte Suprema de Justicia* (Supreme Court of Justice in English) was changed to the *Tribunal Supremo de Justicia* (also translated in English as Supreme Court of Justice), and the Judicial Council, the institution in charge of judicial government, was eliminated. This task was assigned to the "Executive Directorate of Magistrature" under the Supreme Court. |
| 2000 (August 15) | The Supreme Court dictated the "Normative on the Direction, Government, and Management of the Judicial Power," creating a new body under the Supreme Court called the "Judicial Commission." It was composed of six magistrates (one for every chamber) and in charge of judicial |
| | organization and management tasks, including selection and dismissals. |
| 2000 (November 14) | The National Assembly enacted a law establishing that the selection of Supreme Court Magistrates would be by the legislative branch of government; the selection of judges was subject to decisions by qualified majority. At that moment, the National Assembly was formed by a large pro-Chávez majority. |
| 2000-2003 | According to Louza-Scognamiglio (2017), the Supreme Court organized competitive examinations for judicial selection. Despite having around 1,000 vacancies, they opened only 486 positions; from 3,180 candidates, only 270 passed the selection process. In March 2003, the Supreme Court suspended the competitive examinations and dismissed judges without any formal procedure. By the end of 2003, 80% of the judges were provisional. |
| 2003 (October) | The First Administrative Contentious Court was closed, and its magistrates were dismissed by the Judicial Commission. The official reason was serious mistakes and ignorance of the law, but the real cause was that the court issued several resolutions against the government. This was particularly serious because, according to Pérez -Perdomo (2007:19) "this court is legally and politically very important because it has control in the first instance of the legality of the actions of government agencies." |
| 2004 (May 20) | The Organic Law of the Supreme Court was published. According to Louza-Scognamiglio (2017: 105), "This law increased the number of magistrates from 20 to 32, which allowed the National Assembly to appoint pro-government magistrates, and thus, have a majority in the Supreme Court of Justice. This law also established grounds for the removal of magistrates, which gave great discretion to the National Assembly." |
| 2006 | At the end of the year, Chávez proposed a constitutional reform. The president designated a "Presidential Council for Constitutional Reform," chaired by the president of the National Assembly and composed of representatives from |

|  | the rest of the public authorities. "This reform reduced the Venezuelan political and state organization to a single power and changed the political, social, and economic model of the current Constitution. In this type of society, the State owns everything, and there is joint management of public affairs with citizens who can act practically only through the communal councils, which are organs of a new power, the Popular Power, controlled by the Executive Power" (Louza-Scognamiglio, 2017: 106). A constitutional referendum was convened on December 2, 2007, where the constitutional reform was rejected by the majority of votes. |
|---|---|
| 2009 (October 11) | The Judicial System Law was published, creating a new institution, "The National Commission of the Judicial System," to be in charge of the management and government of the judiciary and courts, and the design and enforcement of judicial policies. |
| 2010 (Summer and Fall) | The Organic Law of the Supreme Court was reissued on August 9 and October 1. According to Louza Scognamiglio (2017:110), this was done "in order to speed up the process of appointing new judges before the new National Assembly was installed, since there was a good chance that the government would no longer have full control." |
| 2014 | A new process for selecting magistrates for the Supreme Court was opened. Twelve new magistrates were selected plus four substitutes. Contrary to what was established in the constitution, the new magistrates were selected by a simple majority of the National Assembly (and not by a qualified 2/3 majority) with the intervention of the "Judicial Nominations Committee," made up mostly of members of the ruling party. |
| 2015 (December 6) | Elections were held for the National Assembly. The union of opposition parties (*Mesa de Unidad Democrática*) won the elections and the majority of the seats (112/167). Despite the elections taking place in December 2015, the new National Assembly would be established only later, in January 2016. In the meantime, the outgoing members of the National Assembly (with a solid pro-government |
|  | majority) appointed, in December 2015 by urgent procedure, 13 new magistrates (while inducing early retirements) to guarantee alignment with their political interests. The result was a severe limitation of the newly elected National Assembly's power. |
| 2015-2022 | Several international organizations alerted about the deep and severe problems related to the lack of judicial independence in Venezuela and its effects on human rights. [https://www.icj.org/venezuela-the-authorities-must-stop-undermining-judicial-independence/](https://www.icj.org/venezuela-the-authorities-must-stop-undermining-judicial-independence/) |
| 2022 | A new reform of the Organic Law of the Supreme Court was enacted. The National Assembly elected in 2020 (with a pro-government majority) appointed new Supreme Court magistrates. Despite the constitution allowing a 12-year term for each magistrate, 60% of them were re-elected, violating constitutional principles. (Acceso a la Justicia, 2023:17) |

**Supplementary Appendix A: Counterfactual setup and design**

We observe a series of judicial independence trajectories of $(J \times 1) \in N$ countries in the time period $t = 1, 2, \ldots T$. The constitutional reform takes place at time $T_0 \in (1, T)$ such that $T_0 < T$ and affects only Venezuela which is denoted as $J - 1$-th country and lasts from period $T_0 + 1$ until $T$ without interruption. In addition, let $Q_{j,t}^N$ be the potential trajectory of judicial independence as a realization of the scenario in the eventual absence of the authoritarian assault. Analogically, let $Q_{j,t}^I$ represent the observed trajectory of judicial independence in the presence of the assault. Borrowing the terminology from treatment effects literature, our key parameter of interest is the treatment effect of 1999 constitutional overhaul. Without the loss of generality, the underlying treatment effect can be written as:

$$\gamma_{j,t} = Q_{j,t}^I - Q_{j,t}^N \tag{1}$$

where $Q_{j,t}^I$ is the observed realization of judicial independence indicator in j-th country at time t, and $Q_{j,t}^N$ is the missing counterfactual scenario of judicial independence without the authoritarian assault. Furthermore, the authoritarian overhaul may be described as a binary variable $D$ that can take the value of 1 for the period $t \geq T_0$ and 0 for the period $t < T_0$. Given that Venezuela is the single treated country in our setup with an innate implication that it is indexed as $j = 1$, the observed outcome realization in the presence of authoritarian overhaul is given by:

$$Q_{j,t} = Q_{j,t}^N + \hat{\gamma}_{j,t} \cdot D_{j,t} \tag{2}$$

$$D_{j,t} = \begin{cases} 1 & \text{if } j = 1 \wedge t \geq T_0 \\ 0 & \text{otherwise} \end{cases} \tag{3}$$

Where we estimate the entire vector of post-treatment effects associated with the 1999 authoritarian assault which can be described as the parametric sequence $\{\gamma_{1,T_0+1}, \ldots \gamma_{1,T}\}$. Each element within the vector of post-treatment effects captures the influence of the overhaul on the trajectory of judicial independence from the period $T_0$ until $T$. Per se, $Q_{j,t}^I$ is observable across space and time for the period $t > T_0$ whilst $Q_{j,t}^N$ has to be estimated to uncover the projection of the counterfactual scenario in the hypothetical absence of the overhaul. Henceforth, let $Q_j = [Q_{j,1} \ldots Q_{j,T_0}]$ denote pre-overhaul vector of observed judicial independence trajectories for country $j \in \{1, \ldots J + 1\}$, and let $\mathbf{X_j}$ be a $(K \times 1)$ vector of auxiliary covariates of $\mathbf{Q_j}$. These two distinctive vectors allow us to build $\mathbf{Q_0} = [Q_2 \ldots Q_{J+1}]$ and $\mathbf{Y_0} = [Y_2 \ldots Y_{J+1}]$ matrices invertible across $(K \times J)$ dimension that contain the outcome and covariate values in pre-treatment period for the countries not affected by the authoritarian overhaul at time $T_0$ and beyond.

In the lieu of the binary nature of the treatment, the potential outcome scheme for the judicial independence is given by the following simple latent factor model:

$$Q_{j,t}^{D\in\{0,1\}} = \begin{cases} Q_{j,t}^N(\text{without overhaul} \mid D = 1) = \phi_t + \mathbf{Z}_{j,t}'\theta_t + \lambda_t\mu_j + \varepsilon_{j,t} \\ Q_{j,t}(\text{with overhaul} \mid D = 0) = \quad\quad Q_{j,t}^N + \hat{\gamma}_{j,t} \cdot D_{j,t} \end{cases} \quad (4)$$

where $\phi_t$ represents unobserved time-varying technology shocks to judicial independence common to all countries, $\mathbf{Z}_{j,t}$ is an $(1 \times r)$ vector of observable auxiliary covariates of judicial independence, $\theta_t$ is an $(r \times 1)$ vector of unknown parameters, $\lambda_t$ is an $(1 \times F)$ vector of unknown common factors, and $\mu_j$ is an $(F \times 1)$ vector of unknown factor loadings. The transitory shocks to judicial independence are given by $\varepsilon_{j,t}$ and are assumed to be identically and independently distributed such that $\varepsilon_{j,t} \sim i.i.d$ and $E\left(\varepsilon_{j,t} \mid D(1, T_0)\right) = 0$. It should be noted that the key parameter of interest is $\lambda_t\mu_j$ which allows us to capture the temporal heterogeneity of the response to the constitutional reforms which does not necessitate parallel trend assumption to hold.

Therefore, our goal is to construct a series of artificial control groups for Venezuela to be inasmuch as possible similar to the observed trajectories of judicial independence. More specifically, by reweighing the judicial independence trajectories of the countries using the implicit $\mathbf{Q_0}$ and $\mathbf{Y_0}$ characteristics from the corresponding donor pool in pre-$T_0$ period, the trajectory of $Q_{j,t}^N$ can be estimated for each year in the pre- and post-treatment period denoted as $t \in \{1, \dots T\}$. Under these circumstances, $Q_{j,t}^N$ consists of the reweighted combination of the implicit and explicit attributes of countries from the respective synthetic control group which implies that:

$$\hat{Q}_{j,t}^N = \sum_{j=2}^{J+1} \hat{w}_j Q_{j,t} \quad (5)$$

where $\widehat{\mathbf{W}} = [w_2, \dots w_{J+1}]' \in \mathbb{R}$ describes the vector of weights used to construct the series of synthetic control groups for Venezuela in the hypothetical absence of the judicial overhaul. To approximate the trajectory of the counterfactual scenario, $\widehat{\mathbf{W}}$ can be estimated by finding the solution to the following single-nested minimization problem:

$$\widehat{\mathbf{W}}(\widehat{\mathbf{V}}) = \underset{\mathbf{W} \in \mathcal{W}}{\mathrm{argmin}}(\mathbf{X_1} - \mathbf{X_0}\mathbf{W})'\mathbf{V}(\mathbf{X_1} - \mathbf{X_0}\mathbf{W}) \quad (6)$$

where the vector of weights $W = (w_1, \dots w_j)$ is restricted to be non-negative, convex and additive which implies that $w_j \geq 0$ for each $j \in \{2, \dots J+1\}$ and $\sum_{j=2}^{J+1} w_j = 1$.[14] It

---

[14] It should be noted that the convexity requirement is a sufficient but not a necessary condition to estimate the set of weights used to build the counterfactual trajectory. To fill the void in the literature, Ben Michael et. al. (2021) proposed an augmented version of synthetic control method in settings when pre-treatment quality of the fit is infeasible. More specifically, they propose a bias correction for inexact matching to de-bias the original synthetic control estimate and apply a ridge regression to model the

should be noted that a positive weight computed for $j$-th country from the donor pool suggests that the treated country falls within the convex hull of the implicit outcome- and covariate-specific hull of the $j$-th country where higher weight share indicates a greater degree of similarity therein. Two matrices are built to solve the nested optimization problem and derive the weights to approximate the similarity between Venezuela and the countries from the donor pool. First, **V** represents a diagonal positive semi-definite matrix of $(K \times K)$ dimension with the sum of main diagonal elements that equal one. More precisely, **V** denotes the explanatory power and importance of pre-$T_0$ outcomes and auxiliary covariates in explaining and predicting $Q$ as the outcome of interest. The matrix is constructed and obtained in the training stage where the synthetic control algorithm learns the best model specification to explain the outcome of interest. In the validation stage, the vector of weights **W** is formed denoting the set of countries from the donor pool with an additive structure equal to one, which fall within the identified convex hull of Venezuela's judicial independence attributes in the full pre-treatment period. By default, these two matrices are invertible and ensure that the designated synthetic control group mimics Venezuela's trajectory of outcomes inasmuch as possible. The exact choice of **V** matrix has been a subject of rigorous debate where a variety of solutions has been proposed (Nannicini and Billmeier 2011, Cavallo et. al. 2013, Bohn et. al. 2014, Gobillon and Magnac 2016, Hahn and Shi 2017, Robbins et. al. 2017, Amjad et. al. 2018, Kaul et. al. 2022, Pang et. al. 2022).

Against the backdrop, our approach in the choice of **V** is similar to Abadie et. al. (2015), Klößner et. al. (2018), Firpo and Possebom (2018) and Ferman et. al. (2020). In particular, **V** is chosen through a two-stage procedure. In the initial training period $\widehat{\mathbf{W}}(\widehat{\mathbf{V}}) = \operatorname*{argmin}_{\mathbf{W} \in \mathcal{W}} (\mathbf{X_1} - \mathbf{X_0}\mathbf{W})'\mathbf{V}(\mathbf{X_1} - \mathbf{X_0}\mathbf{W})$ is minimized whereupon we adopt Vanderbei (1999) constrained quadratic optimization routine. This routine is based on a simple algorithm using an interior point method to solve the quadratic programming problem under the imposed constraints. The implementation of the method takes place vis-á-vis C++ plugin where the standard tuning parameters are imposed such as 5% constraint for the tolerance of violation, the maximum number of iterations is set at 1,000 and the clipping bound for the variables is set to 10. In the validation stage, the choice of $\widehat{\mathbf{W}}(\mathbf{V})$ is cross-validated to minimize the out-of-sample prediction error through the follow optimization problem:

$$\widehat{\mathbf{V}} = \operatorname*{argmin}_{\mathbf{V} \in V} (\mathbf{Q_1} - \mathbf{Q_0}\mathbf{W}(\mathbf{V}))'(\mathbf{Q_1} - \mathbf{Q_0}\mathbf{W}(\mathbf{V})) \qquad (7)$$

---

outcomes. Under this approach, the solution to Eq. (6) is bounded on the estimation error which also allows for negative weights and the extrapolation outside the convex hull of the treated unit.

where **V** is a diagonal positive semi-definite invertible matrix of $(K \times K)$ dimension with $\text{tr}(\mathbf{V}) = 1$. More specifically, the matrix of $\mathbf{Q_1}$ is projected on $\mathbf{X_1}$ by imposing $v_k = \frac{|\beta_k|}{\sum_{k=1}^{K}|\beta_k|}$ which denotes the k-the diagonal element of **V** and $\beta_k$ is the k-th coefficient of the linear projection of $\mathbf{Q_1}$ on $\mathbf{X_1}$.

**Supplementary Appendix B: In-Space Placebo Analysis**

Perhaps the most straightforward question arising from estimating the gap between Venezuela as the country affected by the authoritarian assault and its synthetic control group in the follow-up period after the overhaul concerns the statistical significance of the estimated gap. Since synthetic control analysis is a non-parametric technique, statistical significance cannot be conducted using conventional parametric inference and test statistics. Instead, to determine whether the vector of estimated post-reform gaps between Venezuela and its synthetic control group is statistically significant or not, we rely on the treatment permutation procedure that has been proposed in the policy evaluation literature by Abadie et. al. (2010), and has been furthered by Hahn and Shi (2017), Ferman and Pinto (2021), Firpo and Possebom (2018) and Chernozhukov et. al. (2021) among several others.

### B.1   Standard in-space placebo analysis

Under the treatment permutation, the respective policy of interest is assigned to the full set of countries in the donor pool that never underwent the respective policy change in the period of investigation. Once the policy is assigned to other countries, the synthetic control estimator is iteratively applied to each respective country in the donor pool whilst shifting the treated country into the donor pool. Through a series of iterative runs, the distribution of placebo effects is built which can be compared to the estimated gap of the treated country. In brief quantitative terms, treatment permutation and placebo analysis can be described as follows. For each country $j \in \{2, ... J+1\}$ in the period $t \in \{1, ... T\}$, a full vector of post-treatment effects designated as $\hat{\gamma}_j = \{\hat{\gamma}_{1,T_0+1}, ... \hat{\gamma}_T\}$ is estimated upfront. In the subsequent step, the full distribution of placebo effects' vector is built through the permutation of the treatment-related policy to the unaffected countries, and is compared with the empirical distribution of the full treatment effect for Venezuela as a treated country of interest, denoted as $\hat{\gamma}_1 = \{\hat{\gamma}_{1,T_0+1}, ... \hat{\gamma}_{1,T}\}$.

The general *raison d'etre* of such placebo analysis is both simple and straightforward. If the estimated series of judicial independence gaps in response to the constitutional overhaul for Venezuela is like the gaps in the placebo distribution, the notion of significant effect of such reforms would be questionable given that the countries in the donor pool follow similar trajectories of outcomes as Venezuela. In such circumstances, it is likely that a common shock roughly experienced by a large group of countries in the donor pool would preclude the inference of significant effect of constitutional reforms on judicial independence. By contrast, if the estimated judicial independence gap for Venezuela is unique, imperceptible elsewhere in the donor pool, and large, then our analysis may well provide some tentative evidence of the significant effect of the authoritarian assault as such notion then becomes more credible and plausible. The underlying null hypothesis behind the estimated effect of populist constitutional reforms is rejected if the vector of full treatment effects for Venezuela is unusually large and the fraction of countries with similar magnitude of effects and the same sign of the gap is less than in the range between 10 percent and 15 percent.

In comparison with the empirical placebo distribution $|\hat{\gamma}_{j,t}|$, the distribution of $|\hat{\gamma}_{1,t}|$ can be abnormally large in some period after the policy intervention but not in the full period which implies that the exact decision rule behind the rejection of the null hypothesis of zero effect may be difficult. A simple and plausible approach to overcome such innate ambiguity has been advocated by Abadie et. al. (2010) to compare the root mean square prediction error of the treated unit (i.e., Venezuela) with the prediction errors in the placebo simulation before and after the policy change. The rationale behind such comparison is eloquent and straightforward. If the effect of the authoritarian assault is specific to Venezuela as the affected country and is only meagrely perceivable elsewhere, the ratio of Venezuela RMSE and the RMSE of the empirical placebo distribution in the period after the intervention should be exceptionally small. The prediction error ratio can be computed as:

$$RMSE_j = \left(\frac{\sum_{t=T_0+1}^{T}\left(Q_{j,t}-\hat{Q}_{j,t}^N\right)^2}{T-T_0}\right) \div \left(\frac{\sum_{t=T_0}^{T}\left(Q_{j,t}-\hat{Q}_{j,t}^N\right)^2}{T-T_0}\right) \tag{8}$$

where $RMSE_j$ denotes the ratio of mean square prediction errors. The respective ratio can be used to compute non-parametric p-values proposed by Cavallo et. al. (2013) and formally validated by Galiani and Quistorff (2017) which denotes the proportion of units from the donor pool having the RMSE at least as large as the RMSE of the treated unit:

$$\mathbb{P} = \frac{\sum_{j=1}^{J+1} 1\cdot[RMSE_j \geq R_1]}{J+1} \tag{9}$$

where $1[\cdot]$ is a simple Iversionian dichotomous function, $RMSE_j$ is the root mean square prediction error ratio of j-th unit from the donor pool, and $RMSE_1$ is the counterpart root mean square prediction error of the treated unit. Without the loss of generality, the computed p-value cannot be interpreted in the standard parametric framework through the lens of testing sharp null hypothesis. Instead, it may be interpreted as the proportion of units in the donor pool having the estimated effect of the policy change at least as large as the treated unit. Hence, if the proportion of units is high, the null hypothesis of zero effect cannot be rejected. On the other hand, if the proportion is low and within some specified significance threshold such as 0.15, the null hypothesis can be more easily rejected. Henceforth, if the quasi p-value from treatment permutation is sufficiently low, the notion that the estimated gap for the treated unit was obtained either by fluke, chance or at random would not seem plausible and can be refuted. Furthermore, Firpo and Possebom (2018) provide and discuss sufficient conditions that guarantee the formal validity of the inference alongside the size and power of the permutation test.

Figure B.1 presents the p-values associated with the null hypothesis of no treatment effect from the 1999 constitutional overhaul across each judicial independence outcome, based on an in-space placebo analysis using treatment permutation. The vertical axis displays quasi p-values, approximating the proportion of donor countries whose post-1999 outcome gaps are statistically indistinguishable from Venezuela's, as evaluated through RMSE-based comparisons. The horizontal axis reflects the number of years elapsed since the constitutional intervention.

The evidence indicates that p-values remain consistently low, generally hovering around or below the conventional 10% significance threshold. However, the erosion of judicial independence is not uniform across all outcomes. For example, the treatment effects on court-packing-and to a lesser extent, judicial purges-exhibit low p-values for nearly two decades post-reform, but these gradually rise toward the end of the observation window, suggesting a partially transitory effect. In contrast, the decline in Supreme Court independence, judicial constraints on the executive, compliance with the Supreme Court, and judicial accountability is both immediate and persistent, with no indication of convergence by the end of the sample period. Meanwhile, the deterioration in lower-court independence and the escalation of judicial corruption appear more gradual, taking longer to diverge significantly from the placebo distribution.

It is important to note that these p-values primarily capture the relative uniqueness of Venezuela's post-treatment outcome gaps rather than providing formal inference in the classical sense. Nevertheless, when the synthetic control estimator is iteratively applied to each donor country-none of which experienced the institutional rupture caused by the 1999 introduction of the Judicial Emergency Committee-no other country produces a comparably large or persistent gap, nor does any exhibit a similarly high post/pre-$T_0$ RMSE ratio.

Given the excellent pre-treatment fit between Venezuela and its synthetic counterparts, these findings offer compelling empirical support for our core hypotheses. They underscore the exceptional magnitude and endurance of the institutional breakdown triggered by the 1999 constitutional reform-an overhaul that, under the guise of judicial modernization, facilitated one of the most dramatic erosions of judicial independence in the contemporary Ibero-American context.

**Figure B.1**: In-space placebo analysis of the constitutional overhaul in Venezuela, 1960-2021

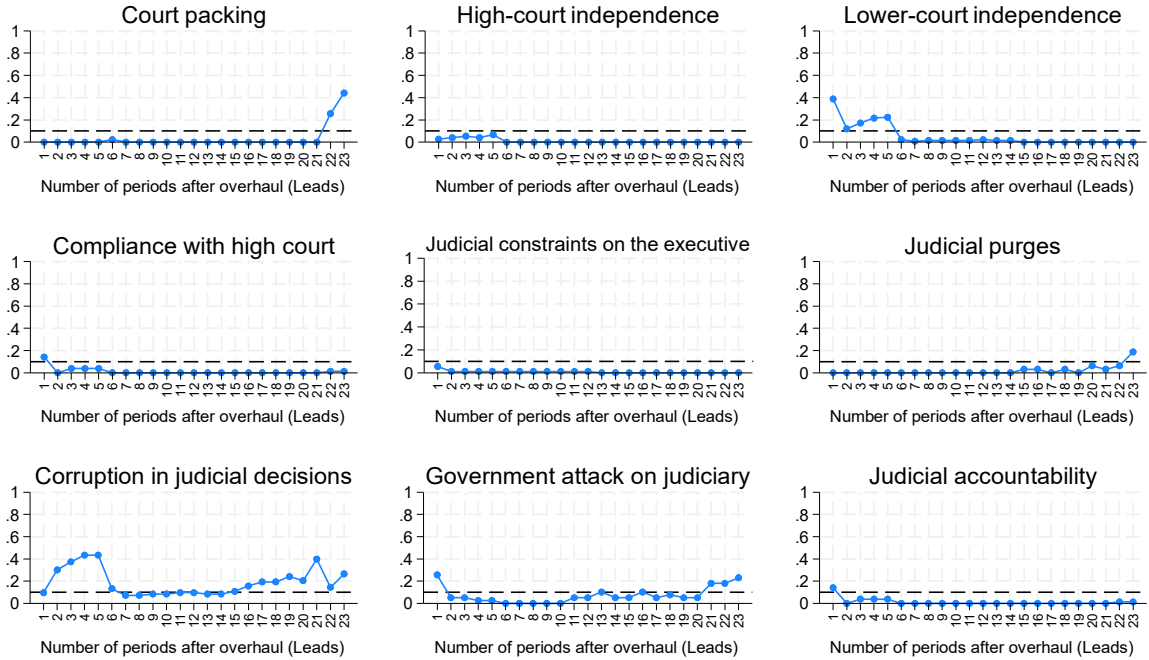

## B.2  Parametric in-space placebo analysis

To obtain a more parametric test and representation of our null outcome-specific hypotheses behind the treatment effect ($\mathbb{H}_0: \hat{\gamma}_{Venezuela, t>T_0} = \{\hat{\gamma}_{1,T_0+1}, ... \hat{\gamma}_T\} = 0$), we employ a difference-in-differences analysis of Venezuela's pre-/post-$T_0$ gaps leveraged against the full distribution of placebo gaps. That said, we estimate a simple difference-in-differences specification where our key parameter of interest is the coefficient on the interaction term between Venezuela and post-$T_0$ period whilst the placebo gaps serve as the control group. The general intuition behind the difference-in-differences approach is simple and straightforward, and complements the non-parametric approach to estimate the quasi p-values behind the null hypothesis from Figure 4. In particular, if the estimated judicial independence gaps are relatively uniquely perceivable to Venezuela and not elsewhere, then the null hypothesis on the interaction term between Venezuela and post-$T_0$ period should be rejected at conventional significance threshold. On the contrary, if the estimated judicial independence gaps are statistically indistinguishable between Venezuela and its placebos, then the null hypothesis could be seldom rejected even at artificially high thresholds. To partially redress the confounding influence of unobserved effects, we add country-fixed and time-fixed effects to each outcome specification, and cap the number of randomly-generated donor samples at 1,000,000 to ensure a steadfast convergence of the random-sampling algorithm across our large-sample iterations.

Table B.1 presents the results of a parametric DiD analysis, estimating Venezuela's treatment effects relative to a large placebo distribution derived from unaffected donor countries. The findings consistently reject the null hypothesis on the interaction term between Venezuela and the post-treatment period at the 1% significance level across all outcomes. The estimated DiD coefficients are uniformly negative and statistically significant, and these effects remain robust in the presence of both country- and year-fixed effects, indicating that they are not artifacts of unobserved heterogeneity or time-specific shocks.

The magnitude of the interaction terms reveals that the most pronounced judicial regressions occurred in the domains of judicial independence, court-packing, compliance with the Supreme Court, and judicial accountability. These patterns suggest that the populist intervention following the 1999 constitutional overhaul entailed a systematic and aggressive assault on the judiciary, characterized by the deliberate erosion of institutional autonomy, the intensification of politically driven judicial appointments, and the breakdown of compliance with the highest court's decisions.

Importantly, the DiD analysis indicates that this authoritarian transformation was not confined to a narrow set of tactics. Even in areas where the treatment effect is relatively less pronounced-such as judicial corruption and constraints on the executive-the estimated DiD interaction terms remain negative and statistically significant at the 1% level. However, their magnitude is approximately one-fourth that observed for the core independence-related outcomes. This asymmetry suggests that while corruption and the weakening of executive constraints played a role, the primary vectors of institutional degradation were the dismantling of judicial independence, the politicization of court composition, and the collapse of judicial accountability mechanisms. Together, these results point to a comprehensive and multifaceted authoritarian assault on the rule of law, with particularly acute effects on the structural integrity of the judicial branch.

Furthermore, we compare the distribution of Venezuela's estimated gaps with the placebo distribution, and compute Glivenko (1933) and Cantelli (1933) empirical p-values on the null hypothesis of Kolmogorov (1933) and Smirnov (1948) equality of distribution functions test, and consecutively reject the null hypothesis at 1%, respectively. The rejection of null hypotheses reiterates further empirical support for the uniqueness and significance of the estimated erosion of judicial independence gaps in response to the 1999 populist assault by judicial emergency committee.

Table B.1: Parametric difference-in-differences analysis of the estimated and in-space placebo judicial independence gaps, 1960-2021

| | Court packing | High-court independence | Lower-court independence | Compliance with high court | Judicial constraints on executive | Judicial purges | Judicial corruption | Government attacks on judiciary | Judicial accountability |
|---|---|---|---|---|---|---|---|---|---|
| | (1) | (2) | (3) | (4) | (5) | (6) | (7) | (8) | (9) |
| Post-$T_0$ $\Delta Y$ | -.434*** | -.424*** | -.404*** | -.307*** | -.084*** | -.315*** | -.068*** | -.287*** | -.437*** |
| | (.038) | (.038) | (.029) | (.039) | (.007) | (.027) | (.006) | (.028) | (.032) |
| Two-tailed 95% confidence bounds | [-.510, -.358] | [-.500, -.347] | [-.462, -.347] | [-.386, -.228] | [-.098, -.069] | [-.370, -.260] | [-.081, -.055] | [-.344, -.230] | [-.502, -.372] |
| Effect persistence (p-value) | [0.000] | [0.000] | [0.000] | [0.000] | [0.000] | [0.000] | [0.000] | [0.000] | [0.000] |
| # observations | 1,116 | 1,116 | 1,116 | 1,116 | 1,116 | 1,116 | 1,116 | 1,116 | 1,116 |
| # countries | 18 | 18 | 18 | 18 | 18 | 18 | 18 | 18 | 18 |
| Within R2 | 0.59 | 0.72 | 0.76 | 0.78 | 0.79 | 0.57 | 0.72 | 0.67 | 0.70 |
| Overall R2 | 0.66 | 0.79 | 0.82 | 0.83 | 0.84 | 0.64 | 0.77 | 0.72 | 0.77 |
| # Random donor samples | 1,000,000 | 1,000,000 | 1,000,000 | 1,000,000 | 1,000,000 | 1,000,000 | 1,000,000 | 1,000,000 | 1,000,000 |
| Permutation method | Random-sampling algorithm | Random-sampling algorithm | Random-sampling algorithm | Random-sampling algorithm | Random-sampling algorithm | Random-sampling algorithm | Random-sampling algorithm | Random-sampling algorithm | Random-sampling algorithm |
| Kolmogorov-Smirnov paired equality of treatment-placebo gap distribution functions (p-value) | 0.000 | 0.000 | 0.000 | 0.000 | 0.000 | 0.000 | 0.000 | 0.000 | 0.000 |
| Country-fixed effects | YES | YES | YES | YES | YES | YES | YES | YES | YES |
| (p-value) | (0.000) | (0.000) | (0.000) | (0.000) | (0.000) | (0.000) | (0.000) | (0.000) | (0.000) |
| Time-fixed effects | YES | YES | YES | YES | YES | YES | YES | YES | YES |
| (p-value) | (0.000) | (0.000) | (0.000) | (0.000) | (0.000) | (0.000) | (0.000) | (0.000) | (0.000) |

Notes: the table reports the post-intervention difference-in-differences coefficients associated with judicial independence gaps after the constitutional overhaul in 1999. In each specification, the full set of country-fixed effects and time-fixed effects is included. Standard errors of the actual and placebo gap coefficients are adjusted for arbitrary heteroscedasticity and serially correlated stochastic disturbances using finite-sample adjustment of the empirical distribution function with the error component model. Cluster-specific standard errors are denoted in the parentheses. Asterisks denote statistically significant coefficients at 10% (*), 5% (**), and 1% (***), respectively.

**Supplementary Appendix C: In-time placebo analysis**

An additional caveat behind the estimated judicial independence gap in response to the populist constitutional reforms emanates from the choice of the treatment year. A pioneering example has been set out by Abadie et. al. (2015) in the study of the economic growth effect of German unification in 1990s on the West Germany drawing on the earlier literature on non-experimental program impact evaluation (Heckman and Hotz 1989). One extant possibility predating the choice of the treatment year to estimate the effect of authoritarian assault concerns the confluence of alternative policies and shocks distinctive from the choice of postulated policy. The general thrust behind these comparisons is relatively simple. Namely, if the estimated erosion of judicial independence in response to the constitutional overhaul is anticipable by distinctive structural breaks taking place in the preceding years, the credibility of the synthetic control estimates and their internal validity can be jeopardized and brought into question (Abadie 2021).

Leveraged against the baseline results, these concerns can be at least partially addressed by conducting an in-time placebo analysis. Contrary to the in-space placebo analysis where policy is assigned to the units in the donor pool, the timing of the constitutional reforms is assigned to a deliberately wrong year by backdating the intervention into the pre-treatment period. Such falsely assigned policy year may be used to gauge temporal placebo analysis and can be seen as falsification test. The confidence in the internal validity of synthetic control estimator would disappear if the method estimated large and similar effects when backdated to the years in which the policy did not take place. By contrast, if a large effect is found for the populist constitutional reforms but no effect is perceptible when the reform period is artificially assigned and backdated to the pre-treatment period, the confidence that the effect estimated for populist constitutional reforms provides a reasonably accurate prediction of trajectories of outcomes for Venezuela becomes more plausible and less susceptible to the alternating influence of distinctive shocks or policies. Additional checks on the choice of the falsely assigned policy year supported by the battery of structural break test are elaborated by Kešeljević and Spruk (2023).

As an additional robustness check aimed at strengthening the internal validity of our estimated judicial independence gaps, unlike the *in-space* placebo tests, which assess whether the estimated effects in Venezuela are unique relative to untreated donor countries, the *in-time* placebo analysis interrogates whether the observed divergence between Venezuela and its synthetic counterpart is truly attributable to the 1999 constitutional overhaul-or, alternatively, to some earlier, unrelated policy shock. This concern can be partially addressed by reassigning the treatment to an intentionally incorrect date prior to the actual reform. However, the standard practice of arbitrarily selecting a placebo year risks introducing interpretive ambiguity and undermining the credibility of the test.

To resolve this, we implement a data-driven approach by applying the Zivot and Andrews (2002) structural break test across the full pre-treatment period (1960–1999). For each judicial independence outcome, we identify the year with the minimum breakpoint test statistic, thus designating it as the most likely candidate for a structural break in the absence of reform. The logic of this strategy is straightforward: if the estimated post-1999 divergence is genuinely linked to the constitutional overhaul, then falsely assigning the treatment to a pre-identified breakpoint should not generate a significant or persistent divergence between Venezuela and its synthetic control.

Figure C.1 presents the estimated structural breakpoints for Venezuela's pre-treatment trajectories across the various judicial independence indicators. Notably, structural breaks are detectable in a few domains-such as court packing (1990), Supreme Court independence (1991), and judicial constraints on the executive (1991)-but are largely absent across the remaining indicators. We therefore proceed by assigning placebo treatments to these breakpoint years, using them as falsification tests. If no comparable divergence is observed in these placebo scenarios, this provides additional empirical support for the claim that the judicial independence gaps observed post-1999 are indeed specific to the constitutional overhaul and not the artifact of earlier institutional fluctuations.

**Figure C.1**: Identifying placebo structural break points in Venezuela's judicial independence trajectories, 1960-2021

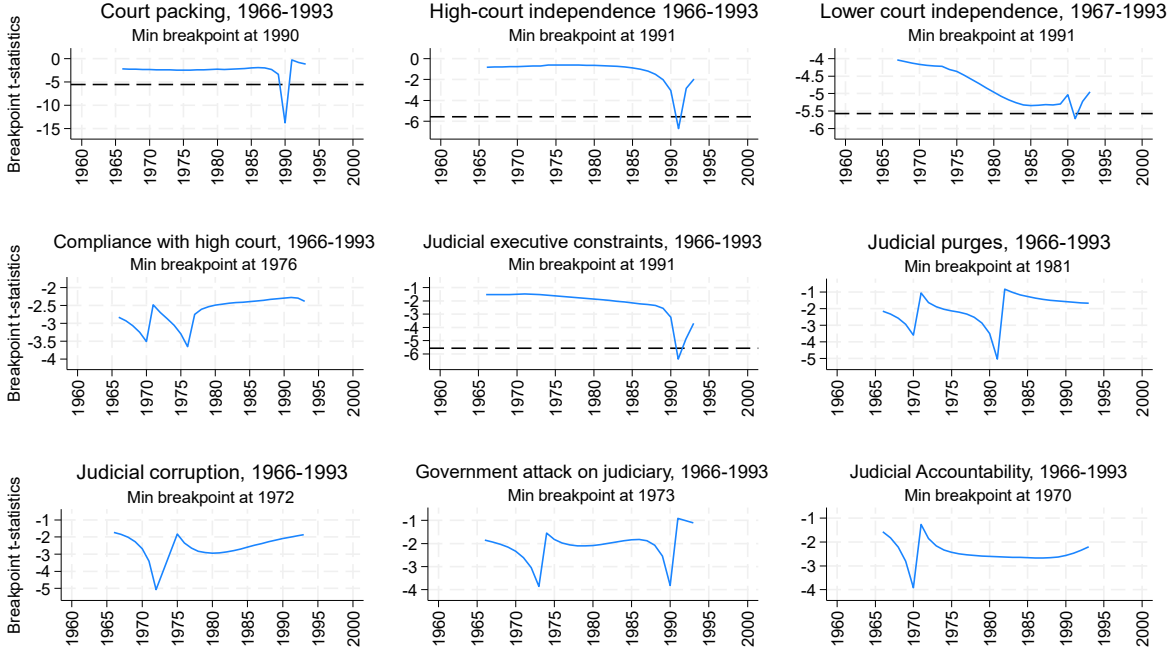

Figure C.2 displays the results of the *in-time* placebo analysis, assessing the effect of falsely assigning the 1999 constitutional overhaul to earlier structural breakpoints in Venezuela's judicial independence trajectories. The vertical dashed line marks the placebo year identified through the Zivot and Andrews structural break test, while the solid vertical line denotes the actual treatment year (1999). As anticipated, the placebo analysis reveals no discernible or systematic divergence between Venezuela and its synthetic counterparts prior to the true reform date. In domains where structural breaks were detected-such as court-packing and Supreme Court independence-the synthetic and actual trajectories remain closely aligned between the placebo year and 1999, with meaningful divergence only emerging after the actual intervention.

This pattern is particularly evident in the case of court-packing, where the treatment effect appears temporary, and in Supreme Court independence, where a near-complete and enduring collapse occurs following the 1999 reform. The absence of pre-treatment divergence reinforces the interpretation that these post-1999 gaps are not the result of earlier institutional shocks or unobserved confounders, but rather a direct consequence of the constitutional overhaul.

Consistent findings are observed across additional outcomes, including lower-court independence, compliance with Supreme Court rulings, judicial constraints on the executive, and judicial accountability. While the trajectories for government-led attacks on the judiciary and judicial corruption exhibit slightly greater noise in the pre-treatment period, the application of the actual 1999 treatment date reveals a persistent and statistically meaningful decline in judicial integrity. Taken together, the in-time placebo results provide strong additional support for the validity of our identification strategy and further affirm that the judicial deterioration observed in Venezuela is not attributable to latent trends or prior policy changes, but to the distinct institutional rupture triggered by the 1999 constitutional reform.

**Figure C.2**: In-time placebo analysis of the constitutional overhaul on Venezuela's judicial independence, 1960-2021

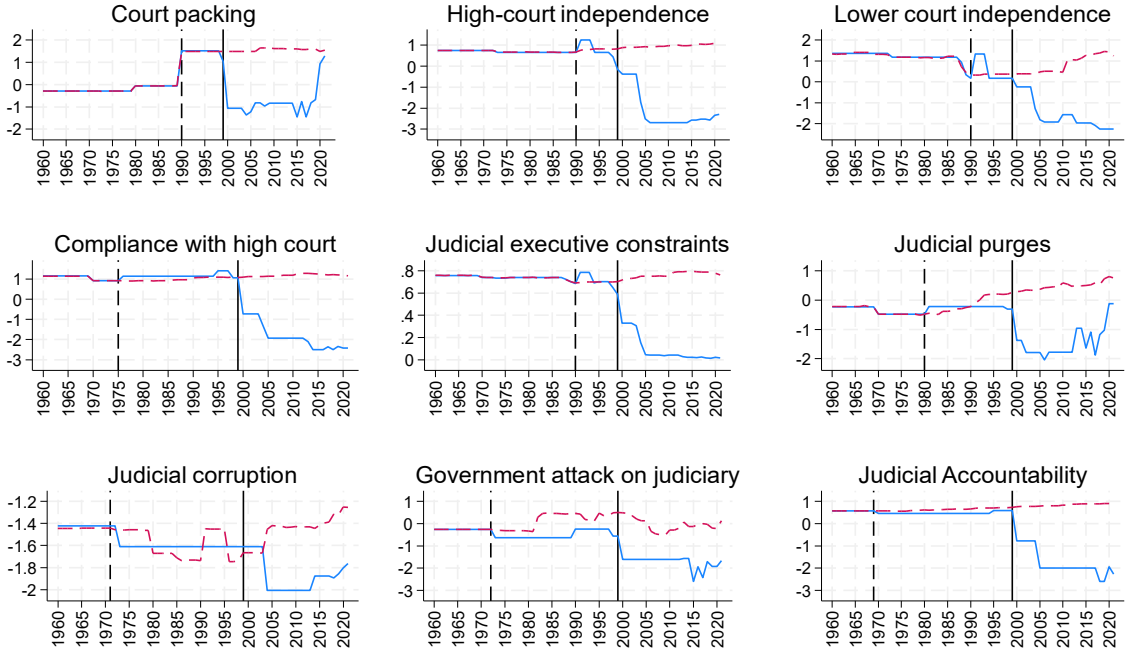

# Supplementary Appendix D: Composition of synthetic control groups

**Figure D.1**: Composition of Venezuela's synthetic control groups

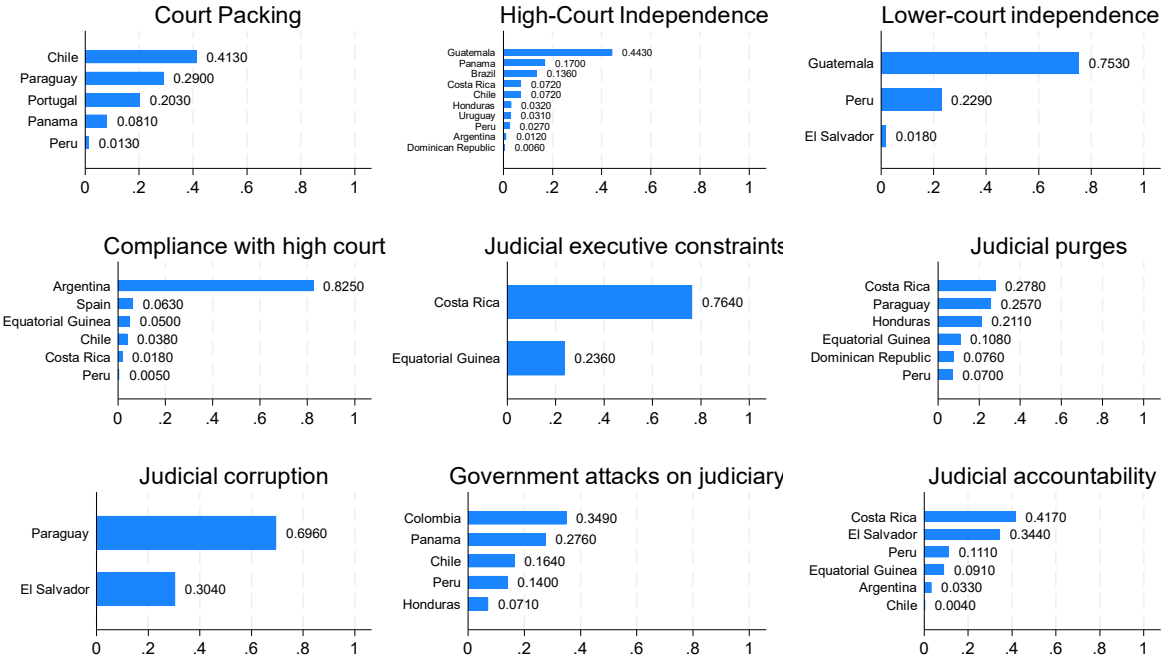

**Supplementary Appendix E: Leave-one-out analysis**

An important consideration for assessing the internal validity of our baseline findings concerns the composition of the synthetic control groups used to approximate Venezuela's judicial independence trajectories. By construction, synthetic counterparts are derived from a weighted combination of countries whose pre-treatment outcome trajectories fall within the convex hull of Venezuela's own pre-overhaul values. These weights are obtained via the convex optimization procedure defined in Equation (6), which identifies the set of donor countries whose outcome profiles best replicate Venezuela's institutional path prior to treatment.

A natural question, then, is whether the estimated effect of the 1999 constitutional overhaul remains robust to variations in the composition of the synthetic control group. To address this, we implement a leave-one-out sensitivity analysis, following Klößner et al. (2018), whereby we iteratively exclude the country contributing the highest weight to the synthetic control in each specification. This exercise enables us to evaluate the extent to which our results are driven by potentially high-leverage donor units.

Figure E.1 presents the leave-one-out replications of the baseline synthetic control estimates, excluding in each case the top-weighted donor country. The results are remarkably consistent with our main findings. The exclusion of any single donor country does not materially alter the estimated post-treatment gap in judicial independence. In particular, the leave-one-out analysis reaffirms a steady and persistent erosion of judicial independence, a breakdown in government compliance with the Supreme Court, and a weakening of both judicial constraints on the executive and mechanisms of accountability. By contrast, the effects on court-packing, judicial purges, and government-led attacks on the judiciary appear somewhat more transitory, though still substantively meaningful.

Overall, the correlation between the baseline estimates and the leave-one-out replications exceeds +0.90 and is statistically significant at the 1% level (p-value = 0.000), indicating a high degree of robustness to changes in donor composition. These findings strengthen the credibility of our causal claims and underscore that the estimated deterioration in Venezuela's judicial institutions is not the artifact of any single influential country in the donor pool, but rather reflects a broader and structurally embedded institutional rupture.

**Figure E.1**: Leave-one-out analysis of the constitutional overhauls effect on Venezuela's judicial independence trajectories, 1960-2021

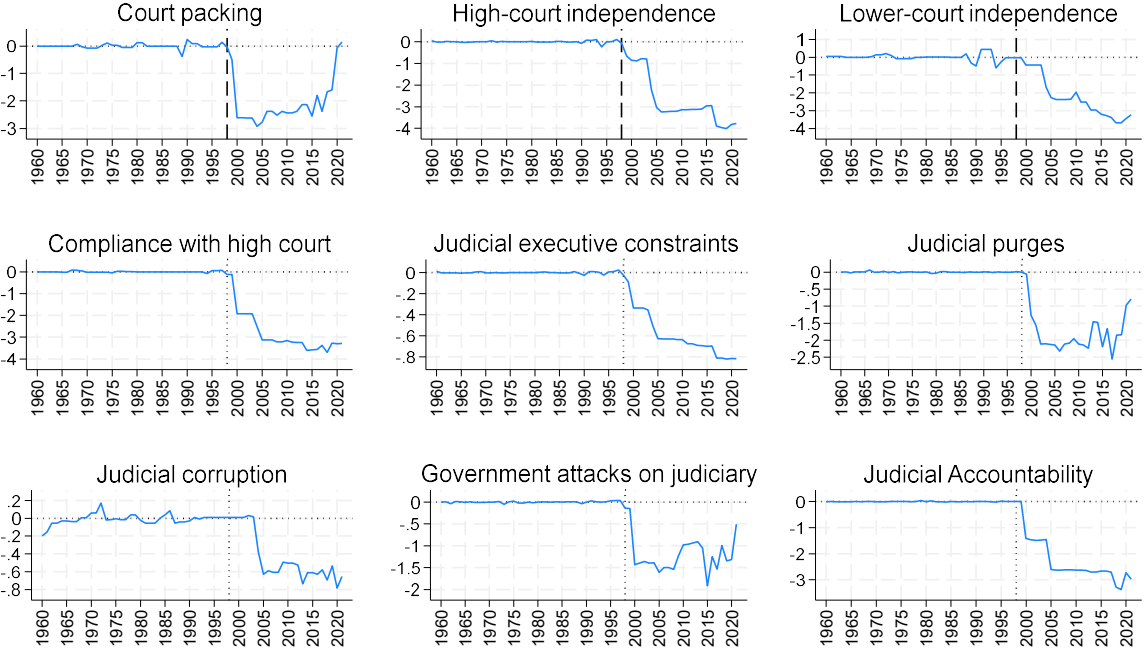

Figure E.2 presents the composition of Venezuela's synthetic control groups under the leave-one-out procedure. The results indicate that even in the absence of the highest-leverage donor country, Venezuela's pre-overhaul judicial independence trajectory can still be effectively replicated by a comparable weighted combination of countries within the convex hull of its pre-treatment characteristics. For example, when Chile-the dominant donor in the baseline court-packing specification-is excluded, Venezuela's pre-overhaul court-packing path is best approximated by a synthetic control composed of Panama (40%), Paraguay (34%), Portugal (20%), Costa Rica (6%), and Brazil (less than 1%).

Similar patterns emerge in other outcomes. The synthetic control groups reproducing Venezuela's pre-reform trajectories for judicial purges, the intensity of government-led attacks on the judiciary, and judicial accountability tend to draw from a broader and more diverse set of Ibero-American donor countries. By contrast, the donor compositions for indicators related to judicial independence, compliance with the Supreme Court, and constraints on executive authority are more compact and parsimonious. These findings further confirm the stability and reliability of our synthetic control estimates, even under alternative donor configurations.

**Figure E.2**: Leave-one-out composition of Venezuela's synthetic control groups

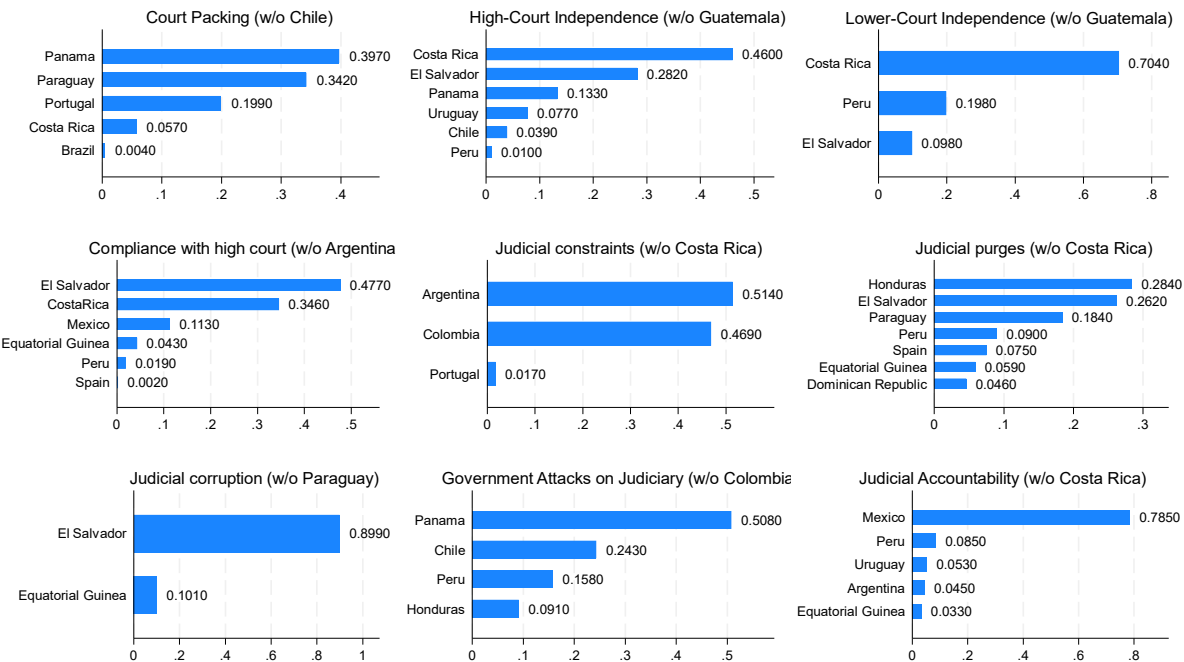

**Supplementary Appendix F: Generalized synthetic control estimates**

One of the key methodological strengths of the synthetic control method lies in its ability to generate counterfactuals without relying on the parallel trends assumption that underpins traditional difference-in-differences designs. Nevertheless, the absence of this assumption introduces its own challenges-particularly when dealing with models subject to unobserved heterogeneity. To address these limitations, Xu (2017) introduces a generalized synthetic control framework that extends standard SCM by estimating counterfactual trajectories under a semi-parametric model structure. In this approach, treatment effects are estimated via a linearly interactive fixed-effects model, wherein time-varying coefficients interact with unit-specific intercepts. Counterfactuals are then constructed through a cross-validation procedure that selects the optimal model specification from the data itself, thereby avoiding arbitrary assumptions, reducing overfitting, and mitigating the curse of dimensionality that often arises from including an excessive number of potentially irrelevant predictors.

In contrast to the classical synthetic control estimator, the generalized approach leverages the full pre-treatment period to construct a reweighting scheme for the control units that optimally approximates the treated unit. To quantify the impact of the 1999 constitutional overhaul, we employ this method to estimate the average treatment effect (ATE) and construct 95% confidence intervals for the post-treatment period, thereby capturing the aggregate uncertainty in the estimated outcome gaps.

Building on this framework, Athey et al. (2021) propose a class of matrix completion estimators that impute missing potential outcomes by recovering the low-rank structure of the outcome matrix in large $N$ and large $T$ settings. These estimators minimize the nuclear norm of the matrix, serving as a convex relaxation of rank minimization, and differ primarily in their choice of regularization parameters. Matrix completion methods have been shown to improve both the consistency and efficiency of counterfactual estimates relative to classical SCM. The regularization parameters are selected through cross-validation: a subset of observed outcomes is used to minimize the average squared prediction error, while ensuring convergence and avoiding overfitting. Following Athey et al.'s recommendation, we implement a fast-convergence algorithm using two variants of the regularization parameter-one corresponding to a low exposure of treatment data to the donor pool, and the other allowing for higher exposure-to test robustness across alternative specifications.

Table F.1 reports the results from the generalized synthetic control framework applied to Venezuela's judicial independence over the 1960–2021 period. Panel A presents estimates derived from standard country- and year-fixed effects. Consistent with prior results, we find that the 1999 populist constitutional overhaul is associated with a statistically significant increase in court-packing, judicial purges, and government-led attacks on the judiciary. The ATE estimates are significant at the 1% level across all outcomes, and the 95% confidence intervals remain narrow, reinforcing the strength and precision of the findings. Additionally, the estimates confirm severe and persistent deterioration in Supreme Court and lower-court independence, the near-total collapse of compliance with high-court rulings, the erosion of judicial constraints on executive authority, and a steady weakening of judicial accountability mechanisms.

Panel B incorporates interactive fixed effects by allowing for interactions between country- and year-specific effects to generate a more flexible and accurate representation of the counterfactual scenario. This refinement improves model fit and sharpens inference. The interactive fixed-effects estimates closely track those in Panel A, with a correlation exceeding +0.90 (p-value = 0.000), and further narrow the confidence intervals due to enhanced precision in pre-treatment matching. Across all specifications, the null hypothesis of zero average treatment effect is rejected at the 1% level, confirming the robustness of the results.

Panel C presents estimates based on matrix completion using a high-regularization hyperparameter to recover the pre-treatment outcome matrix. As expected, the matrix completion results corroborate the generalized synthetic control findings, identifying a pervasive and enduring decline in judicial independence following the 1999 reform. Once again, the most pronounced effects are observed in Supreme Court independence and compliance with judicial rulings. Although the estimated effects on judicial constraints and corruption are of somewhat smaller magnitude, both remain statistically significant at the 1% level. These findings underscore the broader institutional consequences of populist legal engineering and reaffirm the pattern of structural erosion documented throughout our analysis.

Figure F.1 illustrates the average treatment effects and associated 95% confidence intervals obtained from the interactive fixed-effects implementation of the generalized synthetic control method, providing a visual summary of the persistent and statistically significant institutional deterioration triggered by the 1999 constitutional overhaul.

**Table F.1**: Generalized interactive fixed effects and matrix completion estimated effect of constitutional overhauls on Venezuela's judicial independence trajectories, 1960-2021

| | Court packing | High-court independence | Lower-court independence | Compliance with high court | Judicial constraints on executive | Judicial purges | Judicial corruption | Government attacks on judiciary | Judicial accountability |
|---|---|---|---|---|---|---|---|---|---|
| | (1) | (2) | (3) | (4) | (5) | (6) | (7) | (8) | (9) |
| Panel A: Fixed-effects estimates | | | | | | | | | |
| Post-$T_0$ $\Delta Y$ | -1.254*** | -3.298*** | -2.977 | -3.193 | -.725*** | -1.565*** | -.157*** | -1.175*** | -2.418*** |
| | (.066) | (.070) | (.066) | (.062) | (.013) | (.054) | (.060) | (.066) | (.049) |
| Two-tailed 95% confidence bounds | (-1.376, -1.126) | (-3.440, -3.157) | (-3.116, -2.864) | (-3.305, -3.078) | (-.749, -.700) | (-1.683, -1.473) | (-.272, -.042) | (-1.291, -1.020) | (-2.505, -2.325) |
| Simulation-based p-value | 0.000 | 0.000 | 0.000 | 0.000 | 0.000 | 0.000 | 0.009 | 0.000 | 0.000 |
| Panel B: Interactive fixed-effects estimates | | | | | | | | | |
| Post-$T_0$ $\Delta Y$ | -2.461*** | -2.912*** | -1.933*** | -2.969*** | -.595*** | -1.396*** | -.327*** | -1.935*** | -2.386*** |
| | (.424) | (.134) | (.208) | (.086) | (.022) | (.103) | (.046) | (.202) | (.071) |
| Two-tailed 95% confidence bounds | (-3.471, -1.927) | (-3.036, -2.484) | (-2.199, -1.465) | (-3.122, -2.785) | (-.643, -.544) | (-1.618, -1.147) | (-.437, -.253) | (-2.284, -1.444) | (-2.547, -2.261) |
| Simulation-based p-value | 0.000 | 0.000 | 0.000 | 0.000 | 0.000 | 0.000 | 0.000 | 0.000 | 0.000 |
| Panel C: Athey et. al. (2021) fast-convergence matrix completion estimates with high regularization hyper-parameter (lambda = 0.05) | | | | | | | | | |
| Post-$T_0$ $\Delta Y$ | -1.476*** | -3.244*** | -2.823*** | -3.169*** | -.725*** | -1.557*** | -.174*** | -1.203*** | -2.410*** |
| | (.078) | (.056) | (.049) | (.061) | (.013) | (.064) | (.049) | (.058) | (-048) |
| Two-tailed 95% confidence bounds | (-1.610, -1.294) | (-3.358, -3.123) | (-2.935, -2.725) | (-3.301, -3.050) | (-.749, -.699) | (-1.687, -1.433) | (-.262, -.082) | (-1.307, -1.082) | (-2.510, -2.305) |
| Simulation-based p-value | 0.000 | 0.000 | 0.000 | 0.000 | 0.000 | 0.000 | 0.000 | 0.000 | 0.000 |

Notes: the table reports the average treatment effect (ATT) of the populist assault on judicial independence in Venezuela for the period 1960-2021 using interactive fixed-effects and matrix completion estimators. Matrix completion estimator is applied in two distinctive variants of the regularization-based hyper parameter using nuclear space norms. Standard errors are adjusted for serially correlated stochastic disturbances using cluster-robust error component model and finite empirical distribution function, and are reported in the parentheses.

**Figure F.1**: Interactive fixed-effects generalized synthetic control estimated effect of populist assault on judicial independence in Venezuela, 1960-2021

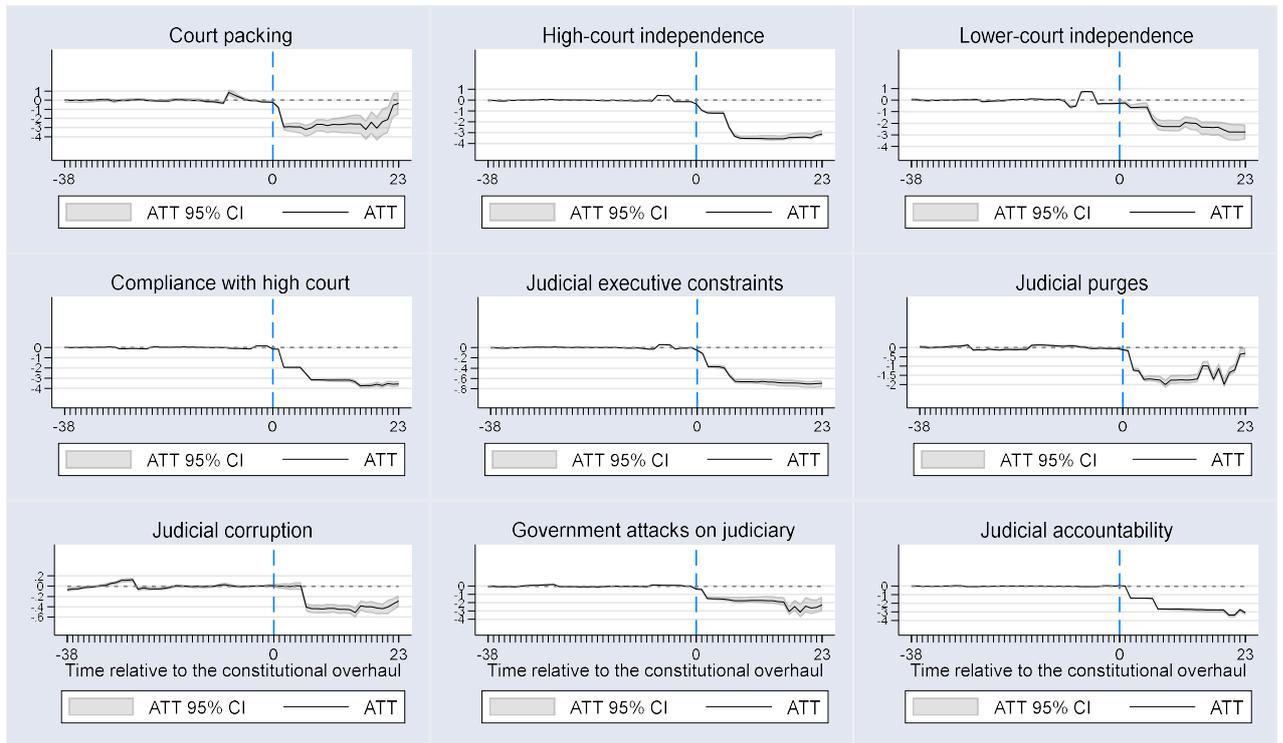

# Supplementary Appendix G: Analysis using Mercosur and OPEC member states as a donor pool

The empirical analysis presented thus far provides robust evidence of a significant erosion of judicial independence in response to the large-scale authoritarian intervention initiated by the 1999 constitutional overhaul and the subsequent rule of a government-appointed judicial emergency committee. A central question that follows concerns the plausibility and credibility of the estimated gaps in judicial independence between Venezuela and its synthetic counterparts. One of the key constraints on the internal validity of these estimates lies in the composition of the donor pool used to construct the synthetic control. Specifically, the inclusion of an overly heterogeneous set of donor countries may introduce interpolation bias, particularly if some units are subject to idiosyncratic shocks or structural breaks that distort their comparability to Venezuela.

As Abadie (2021) emphasizes, the validity of the synthetic control estimator depends critically on its ability to closely replicate the trajectory of the treated unit in the pre-intervention period. Countries experiencing substantial structural volatility or exposed to unobserved shocks may violate the assumptions of the linear factor model-particularly the stability of factor loadings-and should therefore be excluded to preserve estimator credibility. In this context, restricting the donor pool to countries with institutional, historical, and regional similarities enhances the plausibility of the counterfactual.

To mitigate potential interpolation bias and enhance the structural coherence of the control group, we construct a refined donor pool composed exclusively of Mercosur-affiliated countries. This regionally focused comparison set offers greater institutional and geopolitical affinity with Venezuela and reduces the likelihood of confounding shocks that would violate the Stable Unit Treatment Value Assumption (SUTVA). By limiting the donor pool in this manner, we increase the stability and interpretability of the estimated treatment effects.

Figure G.1 replicates the baseline analysis using this Mercosur-based donor pool. The results strongly corroborate the original findings, revealing a gradual yet profound deterioration of judicial independence and accountability in the aftermath of the 1999 reform. The quality of the pre-treatment fit remains high, demonstrating the capacity of the Mercosur[15] donor group to approximate Venezuela's judicial trajectory prior to

---

[15] Mercosur, or the Southern Common Market, is a South American trade bloc established by the Treaty of Asunción (1991) and the Protocol of Ouro Preto (1994). Originally formed by Argentina, Brazil, Paraguay, and Uruguay, it later expanded to include Venezuela and Bolivia. Its primary

the overhaul. The post-treatment comparison yields three salient insights: (i) a rapid but transitory surge in politically motivated judicial appointments; (ii) a pronounced and enduring decline in both Supreme Court and lower-court independence, accompanied by the near-complete collapse of judicial constraints and compliance; and (iii) a long-term degradation of judicial accountability, coupled with an intensified pattern of government-led attacks on the judiciary.

The strength of these findings is further supported by a high correlation between the baseline estimates and those derived from the Mercosur-restricted donor pool. The correlation coefficient exceeds +0.90 and is statistically significant at the 1% level (p-value = 0.000), underscoring the robustness of the results across alternative donor group specifications and reinforcing the conclusion that the 1999 constitutional overhaul constituted a critical rupture in the institutional trajectory of Venezuela's judiciary.

**Figure G.1**: Effects of constitutional overhaul of judicial independence in Venezuela using Mercosur country-level donor pool, 1960-2021

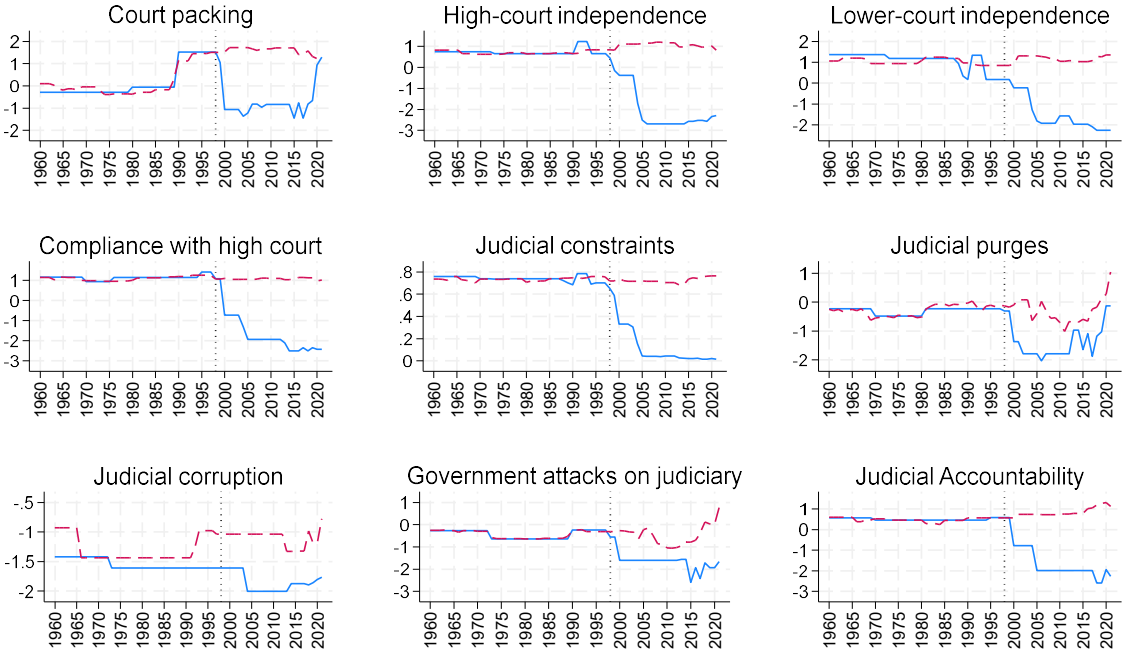

Figure G.2 presents the composition of Venezuela's synthetic control groups constructed from a donor pool of Mercosur countries, selected to best replicate its

---

aim is to foster regional integration and enhance the global competitiveness of Latin American economies. Mercosur has also granted Associated State status to several countries, allowing them to participate in its activities and benefit from trade preferences. Over time, it has signed a broad range of commercial, political, and cooperation agreements with partners across five continents.

judicial independence trajectories prior to the 1999 constitutional overhaul. For example, Venezuela's pre-overhaul trajectory of Supreme Court independence is most closely reproduced by a convex combination of the institutional attributes of Guyana (62%), Suriname (30%), Paraguay (7%), and several others-including Peru, Brazil, and Uruguay-each contributing less than 1%. Similarly, the pre-overhaul trajectory of compliance with the Supreme Court is best approximated by a weighted average of outcome-path characteristics from New Zealand (40%), Guyana (35%), Paraguay (11%), Argentina (6%), Peru (6%), Chile (2%), and Bolivia (less than 1%).

These patterns suggest that a weighted convex combination of Mercosur countries is capable of closely replicating Venezuela's pre-treatment institutional trajectory, with minimal imbalance before the intervention. To further enhance the robustness and interpretive salience of our estimates, we construct two additional donor pool variants, each more restricted in composition but theoretically meaningful.

The first variant draws on member states of the Organization of American States (OAS)[16], enabling a regionally anchored comparison that reinforces the plausibility of the estimated average treatment effect. The second variant is based on a donor pool composed of member countries of the Organization of the Petroleum Exporting Countries (OPEC)[17], offering an alternative benchmark grounded in shared political economy characteristics. Across both specifications, the results yield substantively similar negative effects on judicial independence, supported by strong pre-treatment fit and minimal outcome-path discrepancy prior to the 1999 reform.[18]

---

[16] The Organization of American States (OAS) serves as the principal regional forum for political dialogue, policy coordination, and decision-making in the Western Hemisphere. It is a multilateral institution dedicated to promoting human rights, electoral integrity, social and economic development, and regional security. The OAS comprises 32 member states, including: Antigua and Barbuda, Argentina, the Bahamas, Barbados, Belize, Bolivia, Brazil, Canada, Chile, Colombia, Costa Rica, Cuba, Dominica, the Dominican Republic, Ecuador, El Salvador, Grenada, Guatemala, Guyana, Haiti, Honduras, Jamaica, Mexico, Nicaragua, Panama, Paraguay, Peru, Saint Kitts and Nevis, Saint Lucia, Saint Vincent and the Grenadines, Suriname, Trinidad and Tobago, the United States, Uruguay, and Venezuela.

[17] The Organization of the Petroleum Exporting Countries (OPEC) is an intergovernmental organization that facilitates cooperation among major oil-producing nations with the aim of influencing global oil markets and optimizing revenue from oil production. Established on September 14, 1960, in Baghdad, OPEC currently comprises 12 member states that collectively account for approximately 30% of global oil output. Its membership includes Algeria, the Republic of the Congo, Equatorial Guinea, Gabon, Iran, Iraq, Kuwait, Libya, Nigeria, Saudi Arabia, the United Arab Emirates, and Venezuela.

[18] For the sake of space limitations, additional analyses are not reported but are available upon request.

**Figure G.2**: Composition of Venezuela's synthetic control groups using Mercosur donor pool

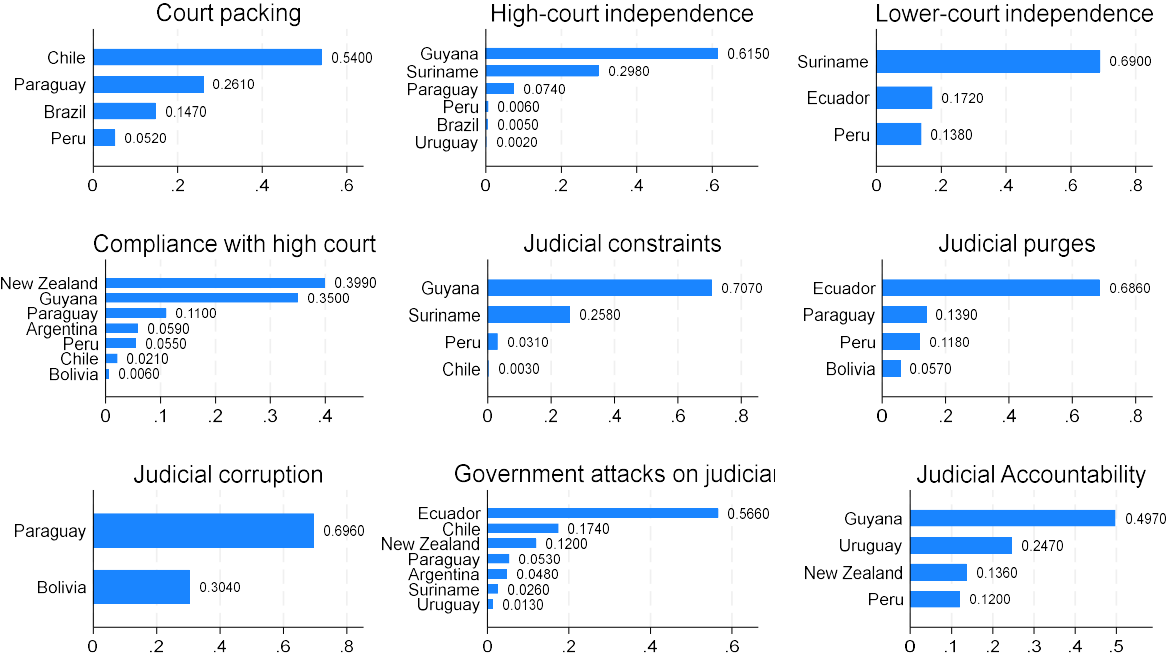

**Supplementary Appendix H: LASSO synthetic control analysis under non-convex optimization**

The synthetic control analyses presented thus far rest on the assumption of common support, which posits that Venezuela's pre-1999 judicial independence trajectory can be accurately reconstructed using a weighted combination of donor countries whose institutional characteristics lie within the convex hull of Venezuela's own. While this assumption ensures a close pre-treatment fit, it imposes a linearity constraint that may be overly restrictive and limit extrapolative flexibility (Ben-Michael et al. 2021). In response to these limitations, Hollingsworth and Wing (2020) propose a more flexible extension of the synthetic control method, employing LASSO-based machine learning regressions. This approach permits extrapolation beyond the convex hull, accommodates both positive and negative weights, and relaxes the unit-sum constraint, allowing counter-cyclical weights that may fall below zero or exceed one.

This LASSO-based estimator offers several advantages: it enhances flexibility in donor pool composition, facilitates automatic model selection for treated and placebo units, and accommodates more complex weighting schemes. Provided the assumption of conditional independence of treatment assignment under potential outcomes holds-and in the absence of pre-intervention structural breaks-the latent factor model underpinning this method can plausibly identify the treatment effect of Venezuela's populist constitutional overhaul on judicial independence.

To satisfy the conditional independence assumption and enhance the salience of the comparison, we restrict the donor pool to a more compact and historically coherent group: the 23 member states of the Organization of Ibero-American States (OIAS).[19] These countries share deep historical, linguistic, and cultural ties, offering a more substantively meaningful comparison set. Accordingly, we replicate our baseline synthetic control estimates using both the flexible LASSO-based estimator and the OIAS-restricted donor pool, thereby allowing extrapolation outside of Venezuela's convex hull while retaining institutional relevance.

Figure H.1 displays the LASSO-based synthetic control estimates of the impact of the 1999 constitutional overhaul on Venezuela's judicial independence, using the OIAS donor pool over the period 1960–2021. The results indicate an excellent pre-treatment

---

[19] Andorra, Argentina, Bolivia, Brazil, Chile, Colombia, Costa Rica, Cuba, Dominican Republic, Ecuador, El Salvador, Equatorial Guinea, Guatemala, Honduras, Mexico, Nicaragua, Panama, Paraguay, Peru, Portugal, Spain, Uruguay, and Venezuela.

fit between Venezuela and its synthetic counterpart, lending credibility to the counterfactual construction. The estimated treatment effects confirm a sharp but temporary increase in politically motivated court-packing and judicial purges. Both forms of institutional interference intensify immediately following the constitutional reform and gradually converge with the synthetic control as the restructuring process stabilizes.

In contrast, the deterioration in Supreme Court independence, compliance with judicial rulings, constraints on executive power, judicial accountability, and corruption levels appears both permanent and cumulative. These effects unfold gradually over time but exhibit sustained divergence from the counterfactual trajectory. Similarly, lower-court independence deteriorates in a sustained manner, although the synthetic control predicts a more moderate erosion in this domain. Additionally, the LASSO estimates corroborate the presence of persistent and statistically significant government attacks on the judiciary, as evidenced by a pronounced and enduring negative gap relative to the synthetic counterpart.

Overall, these findings reaffirm the core conclusions of the baseline analysis while demonstrating the robustness of the results to both methodological variation and donor pool composition. The LASSO-based synthetic control approach not only reinforces the causal inference but also expands the analytical scope by allowing for more flexible counterfactual estimation beyond the confines of convexity-based assumptions.

**Figure H.1**: LASSO-estimated effects of constitutional overhaul of judicial independence in Venezuela using Mercosur country-level donor pool, 1960-2021

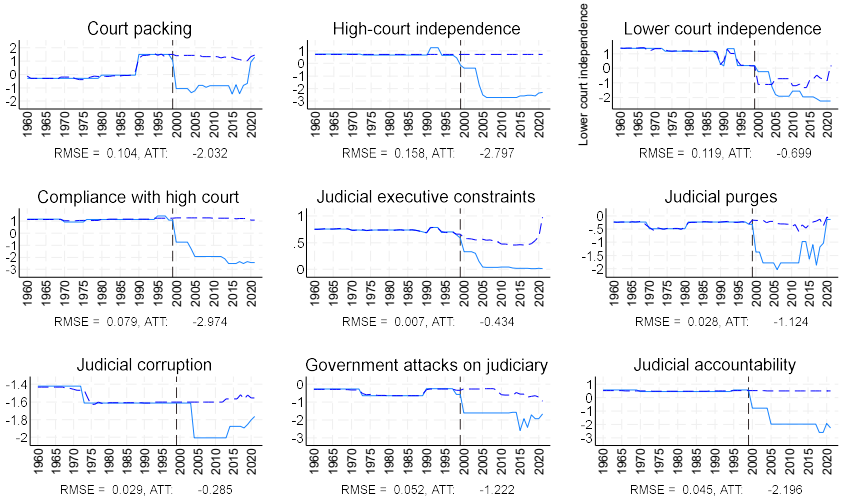

Figure H.2 presents a detailed composition of Venezuela's synthetic control groups estimated using a countercyclical weight structure that allows for extrapolation both within and beyond the convex hull. Under this framework, positive weights indicate similarity in outcome-specific attributes between Venezuela and the donor country, while negative weights denote dissimilarity. The magnitude of the weight, in either direction, reflects the degree of relevance or contrast in reproducing Venezuela's pre-treatment trajectory.

For example, Venezuela's pre-overhaul court-packing trajectory is best replicated through a non-convex combination of institutional characteristics drawn from Chile (38%), Panama (17%), Cuba (13%), Guatemala (11%), Spain (9%), Portugal (7%), Nicaragua (5%), Paraguay (2%), and Argentina (–10%). The negative weight assigned to Argentina suggests its trajectory contrasts with Venezuela's, while the remaining donors contribute positively to the synthetic estimate. A similar pattern is observed in the corruption specification, where the pre-reform trajectory is synthesized by El Salvador (5%), Equatorial Guinea (1%), and negatively weighted contributions from Panama (–1%), Spain (–1%), and Portugal (–19%).

It is important to note that, in this flexible framework, unit-sum and strictly additive weights are not required; weights may vary substantially depending on the complexity of the outcome and the donors' ability to approximate Venezuela's trajectory. Some specifications exhibit broader donor engagement. For instance, the synthetic control group approximating judicial executive constraints includes 20 countries with non-zero weights, while 18 countries contribute to the synthetic reconstruction of the judicial purges trajectory. This expanded and heterogeneous donor composition reflects the structural richness of these institutional dimensions and the flexibility afforded by the non-convex estimation approach.

**Figure H.2**: LASSO-estimated Venezuela's synthetic control groups using a donor pool of Ibero-American states

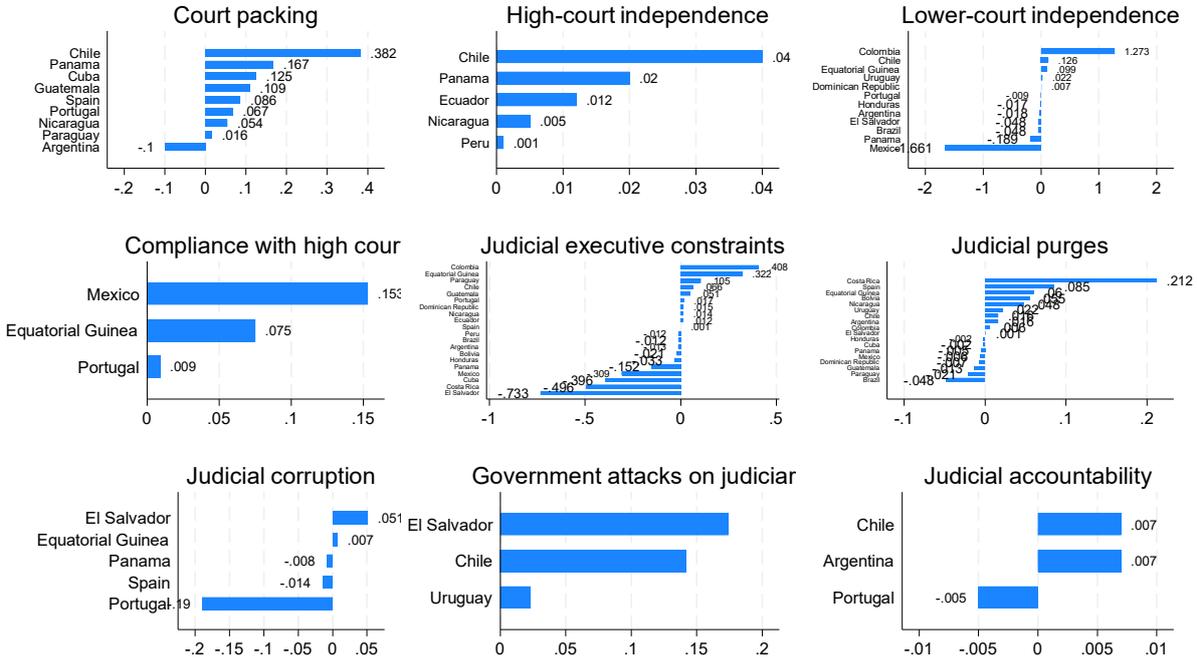

# Supplementary Appendix I: Synthetic difference-in-differences (SDID) analysis

Finally, we consider whether our synthetic control estimates permit stronger causal inference regarding the observed gaps in judicial independence. Causal interpretation beyond mere correlation typically relies on the parallel trends assumption-namely, that in the absence of the 1999 constitutional overhaul, Venezuela would have followed a trajectory of judicial independence similar to that of untreated countries. In practice, this assumption is often difficult to verify and rarely holds in its strict form (Manski and Pepper 2018; Rambachan and Roth 2023). Deviations from the parallel trend undermine point identification, yielding incomplete or partial identification of the treatment effect.

Recent methodological advances have sought to address these limitations by refining the synthetic control framework. These include debiased synthetic control procedures (Ben-Michael et al. 2021) and flexible weighting schemes that permit extrapolation beyond the convex hull (Doudchenko and Imbens 2016). Rather than adopting either the synthetic control estimator, which relaxes the parallel trend assumption, or the traditional difference-in-differences (DiD) estimator, which relies heavily upon it, Arkhangelsky et al. (2021) propose a hybrid approach-synthetic difference-in-differences (SDID)-that combines the strengths of both.

The SDID estimator constructs an optimally matched control group while loosening the dependence on both the parallel trends assumption and the convex hull restriction. It does so by reweighting pre-intervention outcomes using both country-level and time-series weights. This dual weighting framework balances pre-treatment outcome dynamics and improves robustness to additive shifts in the outcome variable. Unlike classical synthetic control, SDID is invariant to affine transformations of the outcome and permits valid inference by combining unit weights with time weights through localized two-way fixed-effects regression. The resulting estimator emphasizes donor countries that exhibit greater similarity to Venezuela in both the pre-treatment and target periods, thereby enhancing consistency and reducing extrapolation bias.

Figure I.1 displays the SDID estimates of the effect of the 1999 constitutional overhaul on Venezuela's judicial independence. The blue line represents the trajectory of the reweighted synthetic control group, derived from the joint country- and year-level weighting scheme; the red line shows Venezuela's actual post-treatment trajectory. The visual and statistical evidence confirms a strong pre-treatment fit, comparable in quality to that achieved by both classical and more flexible synthetic control approaches.

Consistent with our baseline findings and robustness checks, the SDID results reinforce the conclusion that the constitutional reform triggered a systemic and enduring breakdown of judicial independence. The collapse is most clearly visible in the sustained erosion of Supreme Court independence, the near-total dissolution of compliance with judicial rulings, the progressive weakening of constraints on executive authority, and the intensification of judicial purges. Additional effects include a rise in judicial corruption, the increased frequency of formal and informal government attacks on the judiciary, and a profound deterioration in judicial accountability. Collectively, the SDID estimates offer compelling support for a causal interpretation of the post-1999 institutional decline.

**Figure I.1**: Synthetic difference-in-differences estimated effect of populist constitutional overhaul on judicial independence in Venezuela, 1960-2021

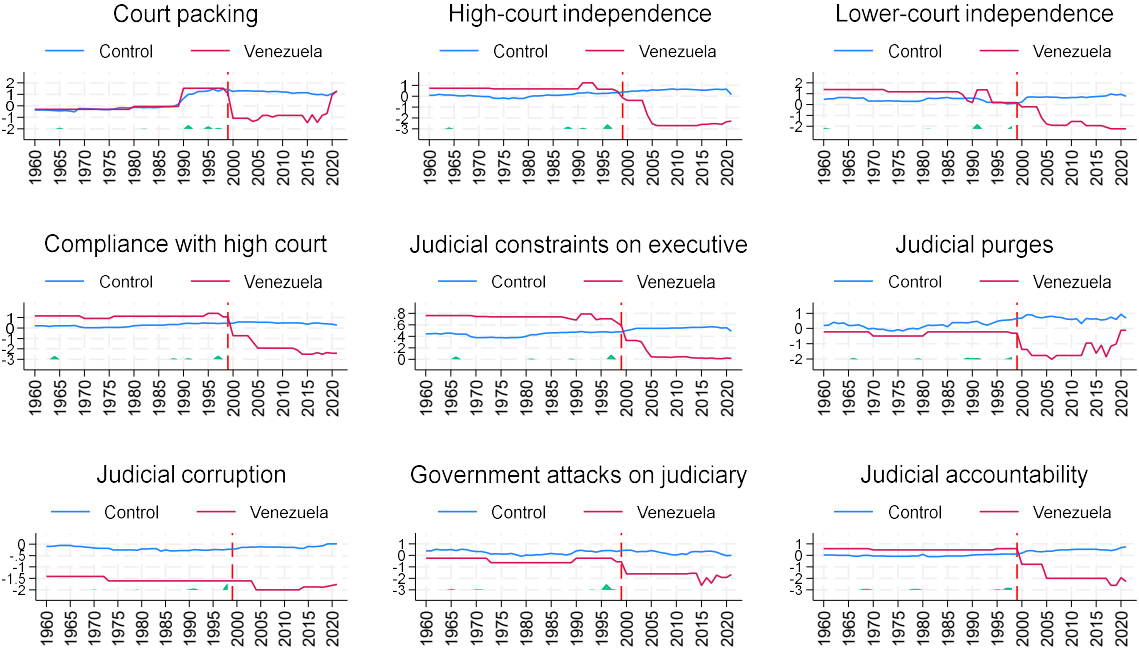

Table I.1 presents the estimated average treatment effects (ATE) of the 1999 constitutional overhaul on each dimension of judicial independence, derived from counterfactual trajectories constructed using a joint country- and year-level weighting scheme. For each outcome, the table reports the point estimate of the ATE, accompanied by two-tailed 95% confidence intervals and empirical p-values based on large-sample approximations under the null hypothesis of no effect.

The results indicate that the most substantial effects are observed in the domains of Supreme Court independence, lower-court independence, and judicial accountability. For eight of the nine judicial outcomes analyzed, the null hypothesis is rejected at the 5% significance level, providing strong evidence of a significant impact. The only exception is judicial corruption, where the p-value (0.119) suggests a marginal or borderline effect.

In addition, the table reports the time-specific weights used to estimate the counterfactual trends, reflecting the degree of temporal similarity between Venezuela's pre-treatment trajectory and those of the control units. Table I.2 complements this analysis by detailing the composition of Venezuela's synthetic control groups, presenting the latent country-level weights derived from the outcome model.

**Table I.1**: Synthetic difference-in-differences estimated effect of 1999 constitutional overhaul on judicial independence in Venezuela, 1960-2021

| | Court packing | High-court independence | Lower-court independence | Compliance with high court | Judicial constraints on executive | Judicial purges | Judicial corruption | Government attacks on judiciary | Judicial accountability |
|---|---|---|---|---|---|---|---|---|---|
| | (1) | (2) | (3) | (4) | (5) | (6) | (7) | (8) | (9) |
| $\tau = \hat{\delta}_{Venezuela} - \sum_{i=1}^{N_{co}} \hat{\omega}_i \hat{\delta}_i$ | -2.220 | -3.135 | -2.797 | -3.140 | -0.694 | -1.412 | -.371 | -1.271 | -2.634 |
| | (0.957) | (0.848) | (.839) | (.956) | (0.179) | (0.741) | (.237) | (.695) | (.465) |
| Two-tailed 95% confidence interval | (-4.096, -.343) | (-4.798, -1.473) | (-4.442, -1.153) | (-5.015, -1.265) | (-1.046, -0.342) | (-2.866, 0.041) | (-.836, .095) | (-2.634, .092) | (-3.547, -1.722) |
| Empirical p-value (large-sample approximation) | 0.021 | 0.000 | 0.001 | 0.001 | 0.000 | 0.057 | 0.119 | 0.068 | 0.000 |
| #non-zero weight donors | 20 | 19 | 11 | 15 | 12 | 19 | 20 | 21 | 17 |
| Time weights: | | | | | | | | | |
| 1960 | | | 0.111 | | | | | | |
| 1961 | | | 0.056 | | | | | | |
| 1964 | | 0.149 | | 0.373 | | | | | |
| 1965 | 0.13 | | | | | | | 0.122 | |
| 1966 | | | | | 0.317 | 0.139 | | | |
| 1967 | 0.025 | | | | | | | | |
| 1968 | | | | | | | | | 0.106 |
| 1969 | | | | | | | | | 0.106 |
| 1970 | | | | | | | 0.048 | 0.071 | |
| 1971 | | | | | | | | 0.043 | |
| 1978 | | | | | | | | | 0.123 |
| 1979 | | | | | | 0.103 | 0.036 | | 0.123 |
| 1981 | | | 0.038 | | 0.093 | | | | |
| 1982 | 0.039 | | | | | | | | |
| 1983 | | | | | <0.01 | | | | |
| 1988 | | 0.238 | | 0.105 | | | | | |
| 1989 | | | | <0.01 | 0.073 | 0.177 | | | |
| 1990 | | | | | | 0.104 | 0.069 | | |
| 1991 | 0.409 | 0.114 | 0.520 | 0.114 | | 0.149 | 0.159 | | |
| 1993 | | | | | | | | 0.042 | |
| 1995 | 0.271 | | | | | | | | 0.081 |
| 1996 | | 0.502 | | | | | | 0.623 | |
| 1997 | 0.126 | | | | 0.511 | 0.074 | | 0.097 | 0.232 |
| 1998 | | | 0.272 | | | 0.252 | 0.686 | | 0.226 |

Notes: the table reports synthetic difference-in-differences estimated effect of the populist constitutional overhaul on judicial independence by matching Venezuela's pre-overhaul judicial independence trajectories with a donor pool of 22 members of the Organization of Ibero-American States. The table reports the weighted difference between Venezuela's judicial independence trajectory and its control group based on the localized two-way fixed effect estimator using outcome model includes latent country-level factors interacted with latent time factors. It also reports latent time-varying weights and large sample-approximated empirical p-value on the null hypothesis behind the average treatment effect. The lower and upper bound of the two-tailed 95% confidence interval is reported in the parentheses.

**Table I.2**: Distribution of time-invariant country-specific weights in the composition of synthetic version of Venezuela

| | Court packing | High-court independence | Lower-court independence | Compliance with high court | Judicial constraints on executive | Judicial purges | Judicial corruption | Government attacks on judiciary | Judicial accountability |
|---|---|---|---|---|---|---|---|---|---|
| Argentina | 0 | 0.055 | 0.069 | 0.110 | 0.064 | 0.030 | 0.091 | 0.036 | 0.029 |
| Bolivia | 0.018 | <0.01 | 0 | 0 | 0 | 0.028 | 0.064 | 0.044 | 0 |
| Brazil | 0.038 | 0.036 | 0 | 0 | 0 | 0.053 | 0.021 | 0.060 | 0.056 |
| Chile | 0.172 | 0.051 | 0 | 0 | 0 | <0.01 | 0.064 | 0.086 | <0.01 |
| Colombia | 0.018 | 0.030 | 0.069 | 0.062 | 0.043 | 0.067 | 0.041 | 0.055 | 0.049 |
| Costa Rica | <0.011 | 0.088 | 0.110 | 0.066 | 0.168 | 0.098 | 0.038 | 0.052 | 0.094 |
| Cuba | 0.143 | 0.095 | 0.110 | 0.106 | 0.160 | 0.029 | 0.060 | 0.051 | 0.094 |
| Dominican Republic | <0.01 | <0.01 | 0 | 0.004 | 0.019 | 0.048 | 0.054 | 0.033 | 0.051 |
| Ecuador | <0.01 | 0.057 | 0.111 | 0.071 | 0.016 | 0.089 | 0.069 | 0.052 | 0.094 |
| Spain | 0.059 | 0 | 0 | 0 | 0 | 0.066 | <0.01 | 0.050 | 0 |
| Equatorial Guinea | 0.032 | 0.012 | 0.012 | 0.072 | 0.155 | 0.055 | 0.024 | 0.025 | 0.074 |
| Guatemala | 0.025 | 0.127 | 0.175 | 0.091 | 0.118 | 0 | 0.020 | 0.040 | 0.015 |
| Honduras | 0.035 | 0.060 | 0 | 0.089 | 0 | 0.085 | 0.060 | 0.056 | 0.047 |
| Mexico | 0.016 | 0.023 | 0 | 0.081 | 0.042 | 0.051 | 0.060 | 0.047 | 0.086 |
| Nicaragua | 0.093 | 0.057 | 0 | 0 | 0 | 0.032 | 0.022 | 0.055 | 0 |
| Panama | 0.128 | 0.099 | 0.050 | 0.053 | 0.055 | 0.054 | 0.029 | 0.051 | 0.023 |
| Peru | 0.004 | 0.018 | 0.169 | 0.036 | 0.055 | 0.057 | 0.108 | 0.016 | 0.143 |
| Portugal | 0.073 | 0 | 0 | 0.019 | 0 | 0 | 0 | 0.052 | <0.01 |
| Paraguay | 0.112 | 0.032 | 0.063 | 0.034 | 0 | 0.026 | 0.032 | 0.042 | 0 |
| El Salvador | 0.012 | 0.090 | 0.062 | 0.106 | 0.105 | 0.078 | 0.065 | 0.053 | 0.086 |
| Uruguay | 0.005 | 0.063 | 0 | 0 | 0 | 0.046 | 0.075 | 0.044 | 0.047 |